\documentclass[preprint2]{aastex}

\newfont{\boldit}{cmbxti10}
\DeclareMathAlphabet{\mitbf}{OML}{cmm}{b}{it}

\usepackage{subfig}
\usepackage{amsmath}

\begin{document}


\title{Structure and evolution of pre-main sequence circumstellar disks}


\author{Andrea Isella, John M. Carpenter and Anneila I. Sargent}
\affil{Department of Astronomy, California Institute of Technology, MC 249-17, Pasadena, CA 91125.}
\email{isella@astro.caltech.edu}


\begin{abstract}
We present new sub-arcsecond (0.7'') Combined Array for Research in 
Millimeter-wave Astronomy (CARMA) observations of the 1.3 mm continuum 
emission from circumstellar disks around 11 low and intermediate mass pre-main 
sequence stars. High resolution observations for 3 additional sources were obtained 
from literature. In all cases the disk emission is spatially resolved. 
We adopt a self consistent accretion disk model based on the similarity
solution for the disk surface density and constrain the 
dust radial density distribution on spatial scales of about 40 AU. 
Disk surface densities appear to be correlated with the stellar 
ages where the characteristic disk radius increases from ~ 20 AU 
to 100 AU over about 5 Myr. This disk expansion is accompanied by
a decrease in the mass accretion rate, suggesting that our sample disks form 
an evolutionary sequence. Interpreting our results in terms of the 
temporal evolution of a viscous $\alpha$-disk, we estimate (i) 
that at the beginning of the disk evolution about 60\% of the circumstellar 
material  was located inside radii of 25--40 AU, (ii) that disks formed with 
masses from 0.05 to 0.4 M$_{\sun}$ and (iii) that the viscous timescale at the 
disk initial radius is about 0.1-0.3 Myr.
Viscous disk models tightly link the surface density $\Sigma(R)$
with the radial profile of the disk viscosity $\nu(R) \propto R^{\gamma}$.
We find values of $\gamma$ ranging from -0.8 to 0.8, suggesting that
the viscosity dependence on the orbital radius can be very different in 
the observed disks. Adopting the $\alpha$ parameterization for the viscosity, we
argue that $\alpha$ must decrease with the orbital radius and that it may
vary between 0.5 and $10^{-4}$. From the inferred disk initial radii we
derive specific angular momenta, $j$, for parent cores of 
$(0.8 - 4) \times 10^{-4}$ km/s pc. Comparison with the
values of $j$ in dense cores suggests that about 10\% of core
angular momentum and 30\% of the core mass are conserved in the formation of the
star/disk system. We demonstrate that the similarity solution for the surface 
density for $\gamma <0$ can explain the properties of some  ``transitional disks'' 
without requiring discontinuities in the disk surface density. In the case of 
LkCa~15, a smooth distribution of material from few stellar radii to
about 240 AU can produce both the observed SED and the spatially resolved continuum
emission at millimeter wavelengths. Finally we show that among the observed 
sample, TW~Hya is the only object that has a disk radius comparable with 
the early solar nebula.

\end{abstract}






\section{Introduction}
Spatially unresolved observations of the infrared and mm-wave 
emission from nearby pre-main sequence stars surrounded by disks 
suggest that most of the circumstellar dust dissipates  
on timescales between 1 and 10 Myr \citep[see, e.g.,][]{hz07}. 
Nevertheless, it remains very uncertain how disk 
evolution proceeds in individual systems and, in particular, whether all circumstellar disks 
give rise to planetary systems. Over the last ten years a large number of 
circumstellar disks in nearby star forming regions have been observed 
using long baseline millimeter and sub-millimeter interferometers. 
These observations have spatially resolved the disk emission to infer 
the radial distribution of gas and dust.
However, at the distance of the nearby star forming regions (100-200 pc),
the typical angular resolution of 1.5\arcsec\--3\arcsec\ 
\citep{rod06,and07,kit02,d96} could not constrain the detailed structure of 
disks, which typically have radii of only a few hundred AU. 

Higher angular resolution mm-wave observations remain challenging and  
the number of disks observed at angular resolution higher than 1\arcsec\ 
is still very small and essentially restricted to more massive and 
luminous pre-main sequence circumstellar disks such as LkH$\alpha$~330 
\citep{b08}, HD~163296 \citep{is07}, AB~Aur, MWC~480, DM~Tau and 
LkCa~15 \citep{pi05,pi06,pi07}, CQ~Tau \citep{t03}, 
DL~Tau, UZ~Tau, BP~Tau and GM~Aur \citep{sim00}, 
TW~Hya \citep{wi00} and GG~Tau \citep{gu99}. 
Even for this small sample the radial distribution and kinematics of the 
circumstellar material vary considerably from object to object.
Since the observed objects are characterized by stellar ages 
between $\sim$0.1 and $\sim$10~Myr, which is probably a considerable 
fraction of the disk life time, variations in the dust properties may 
also be representative of different evolutionary stages. 
Differences in disk structure are believed to result from variation in the
total angular momenta, masses, chemical compositions and magnetic fields, during 
the collapse of the parent molecular core \citep[see, e.g.,][]{hg05}. 
Disk structure in multiple systems or in dense star forming regions 
can also be influenced by the dynamical perturbation induced by close-by 
companions or by strong interstellar radiation field \citep[see, e.g,][]{al06}.
Detailed investigations of disk structure and of the 
origins of any observed difference are clearly necessary to improve our 
understanding of the formation of planetary systems. 
 
Here we present sub-arcsecond observations of circumstellar disks 
around 14 nearby pre-main sequence stars. New 1.3 mm continuum observations 
of 11 objects in the Taurus and Ophiuchus star forming regions, CY~Tau, DG~Tau, 
DM~Tau, DN~Tau, DR~Tau, GO~Tau, LkCa15, RY~Tau, UZ~Tau~E, GSS~39, SR~24~S, 
were obtained with the Combined Array for Research in Millimeter 
Astronomy\footnote{Support for CARMA 
construction was derived from the Gordon and Betty Moore Foundation, the 
Kenneth T. and Eileen L. Norris Foundation, the Associates of the 
California Institute of Technology, the states of California, Illinois, 
 and Maryland,and the National Science Foundation. Ongoing CARMA 
development and operations are supported by the National Science 
Foundation under a cooperative agreement, and by the CARMA partner 
universities} (CARMA). For MWC~275 (HD163296), 
GM~Aur, and TW~Hya, we have reanalyzed published data from the SMA and Plateau de 
Bure interferometers \citep{hu08,is07}. For each object we derive 
the radial dust distribution by comparing the observed dust continuum 
emission with a self-consistent disk model based on the similarity
solution for the surface density of a viscous keplerian disk \citep{lb74}.

In \S~\ref{sec:sample} we summarize the properties of the stellar 
sample. The interferometric observations and data reduction procedures are 
described in \S~\ref{sec:obs}. In \S~\ref{sec:obs_res} we discuss 
the observations. \S~\ref{sec:desc} contains 
the description of the adopted disk model, while the results are presented 
in \S~\ref{sec:res}. The discussion and the conclusions follow in 
\S~\ref{sec:disc} and \ref{sec:conc}.

\section{The sample}
\label{sec:sample}
The 14 stars selected for study are listed in Table~\ref{tab:sample}
together with the adopted luminosities, spectral types, temperatures and
accretion luminosities. All are nearby pre-main sequence T Tauri stars 
with known 1.3~mm flux densities in excess of 50~mJy, to ensure high 
signal-to-noise ratios for the extended disk emission. Two targets, UZ~Tau~E 
and SR24~S, are members of multiple systems. The first nine objects 
of Table~\ref{tab:sample} are located in the Taurus-Auriga star forming region,
while GSS~39 and SR~24~S are in Ophiuchus. TW~Hya is in the homonymous 
association, and MWC~275 (HD~163296) is an isolated Herbig Ae star. 
Based on observations of the molecular gas emission, the strong excess 
continuum emission at IR, mm and radio wavelengths appears to originate from 
large amounts of gas and dust distributed in a 
rotating disk \citep{gu99,d96,sim00,qi04,is07}.

\subsection{Stellar properties}
\label{sec:dist}
For the objects in Taurus-Auriga and Ophiuchus we 
assume stellar distances of 140$\pm$10 and 130$\pm$15 pc respectively 
\citep[see][]{reb04}. For TW~Hya and MWC~275 
we adopt the {\it Hipparcos} distances of 56$\pm$5 pc (Wichmann et 
al.~1998) and 122$\pm$20 pc (van den Ancker et al.~1998) respectively.

Stellar ages and masses are derived
from the H-R diagram using theoretical tracks from \citet[][with the 1998 updated 
version available on the web; hereafter DM97]{dm97} adopting published 
spectral types and luminosities (see Table~\ref{tab:sample}). Assuming 
errors of about 30\% in the stellar luminosity and half a 
spectral type in the spectral type classification, the resulting masses and 
ages are uncertain by 30-50\%. 
Although masses and ages are strongly dependent on the adopted
stellar evolution model (Appendix~\ref{sec:HR}), the main results of the paper
are almost independent of this choice, as discusses in \S~\ref{sec:disc}.


Stellar radii based on the effective temperature and the 
bolometric luminosity were combined with the derived masses to provide 
mass accretion rates $\Dot{M}_{acc}$, following the relation 
$\Dot{M}_{acc}=L_{acc}R_{\star}/(GM_{\star})$. Both quantities are 
listed in Table~\ref{tab:sample}, together with stellar masses and ages.



\section{Observations}
\label{sec:obs}
Interferometric observations of the 11 disks in Table~\ref{tab:obs_1mm} 
were carried out with CARMA, which consists of six 10 m and 
nine 6 m antennas, and is located near Big Pine (CA) at an altitude of about 2200 m.

The data were obtained between Oct 2007 and Apr 2008 using the C and B 
array configurations to provide baseline lengths between 20 and 270~m 
and between 90 and 900~m respectively, corresponding to angular 
resolutions of about 0.7\arcsec\ and 0.4\arcsec\ at 1.3 mm. The CARMA 
correlator was configured with two wide bands of 500~MHz each and one 
narrow band of 8~MHz centered at the frequency of 230.538~GHz, 
corresponding to the $^{12}$CO (2-1) molecular transition.
Only continuum emission observed in the wide 1~GHz band will be
discussed here. Each source was observed for 4-8 hrs to 
achieve a noise level between 1 and 5 mJy. Source names, 
array configurations, synthesized beam sizes, dates and adopted gain 
calibrators are summarized in Table~\ref{tab:obs_1mm}. 

\begin{figure*}
\centering
\includegraphics[angle=0,width=\textwidth]{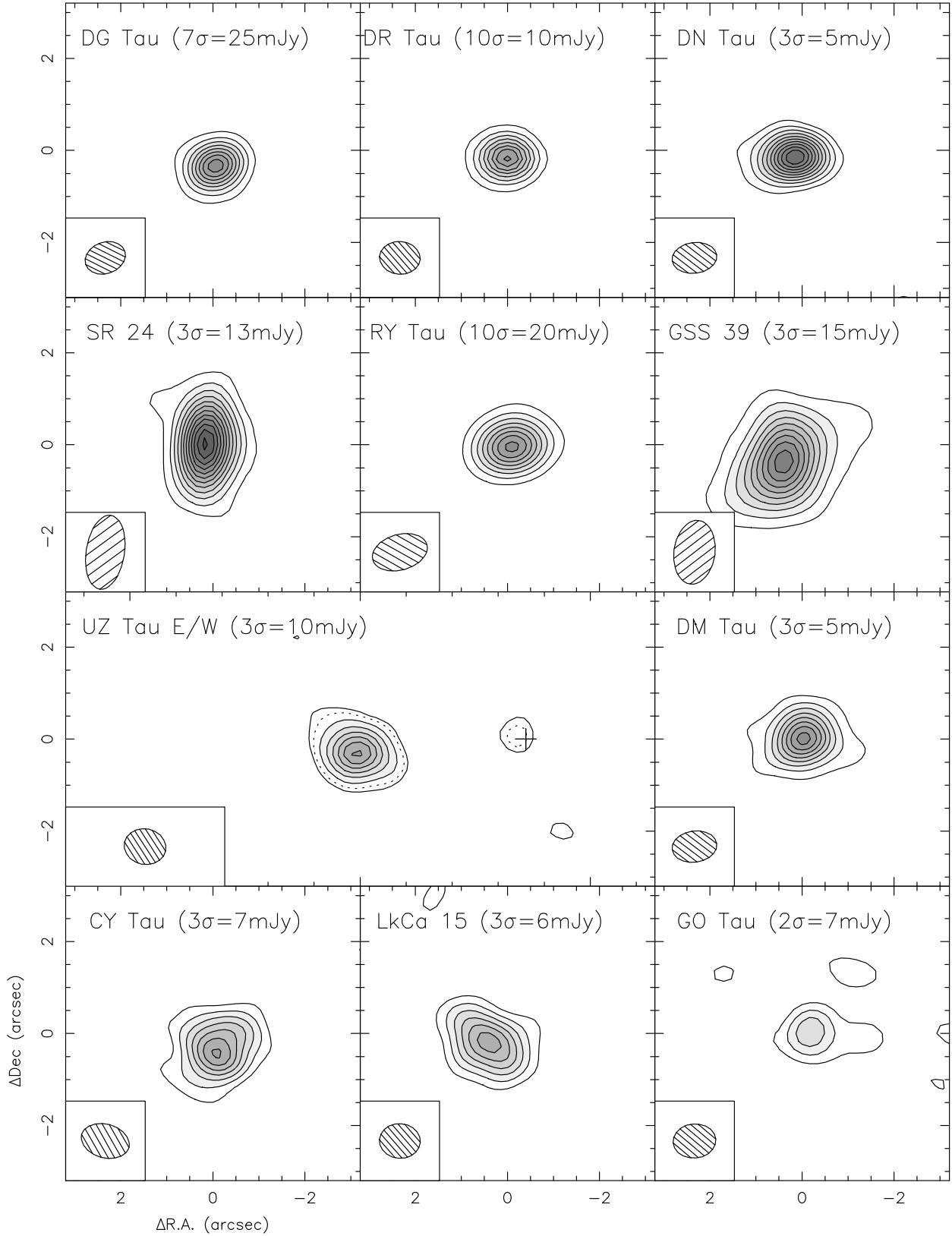}
\caption{\label{fig:cont} 1.3~mm dust continuum images of the disks observed with 
CARMA. Contours start at the significance levels given in each panel and 
are separated by that same amount. The exception is UZ~Tau~E/W where a cross indicates 
the position of UZ~Tau~W and the dotted contour corresponds to the 4$\sigma$ level.
Beam sizes and PA are listed in Table~\ref{tab:obs_1mm}. 
Integrated fluxes and source sizes are given in Table~\ref{tab:gau_fit}.
}
\end{figure*}

The data were reduced using the MIRIAD software package. The 
resulting 1.3~mm maps, obtained using natural weighting, are shown in 
Figure~\ref{fig:cont}. Bandpass calibration relied on observations 
of 3C273 and absolute flux calibration was set by observing Uranus. 
Since we are interested in resolving the disk extended structures we 
took particular care in correcting for the atmospheric turbulence which 
may result in artificially extended sources. For each set of observations, 
we calibrated the data by observing every 15 minutes a bright unresolved 
calibrator located within 20\arcdeg\ from the target. Fitting a bidimensional 
Gaussian profile to the calibrator image we measured the seeing and 
corrected the data using the SEEING 
option of the UVCAL MIRIAD command. Residual effects of the seeing are 
then eliminated by rejecting the UV points on which the atmospheric 
turbulence introduces a flux loss higher than 10\% on the phase calibrator. 
To check the quality of the phase calibration we observed, if available, 
a fainter second point source located within 5\arcdeg\ from to the science 
target. We then verified that, after applying the calibration solution 
derived for the bright calibrator, we obtain an unresolved image of this 
second calibrator.
  
\begin{figure*}
\centering
\includegraphics[angle=-90,width=\textwidth]{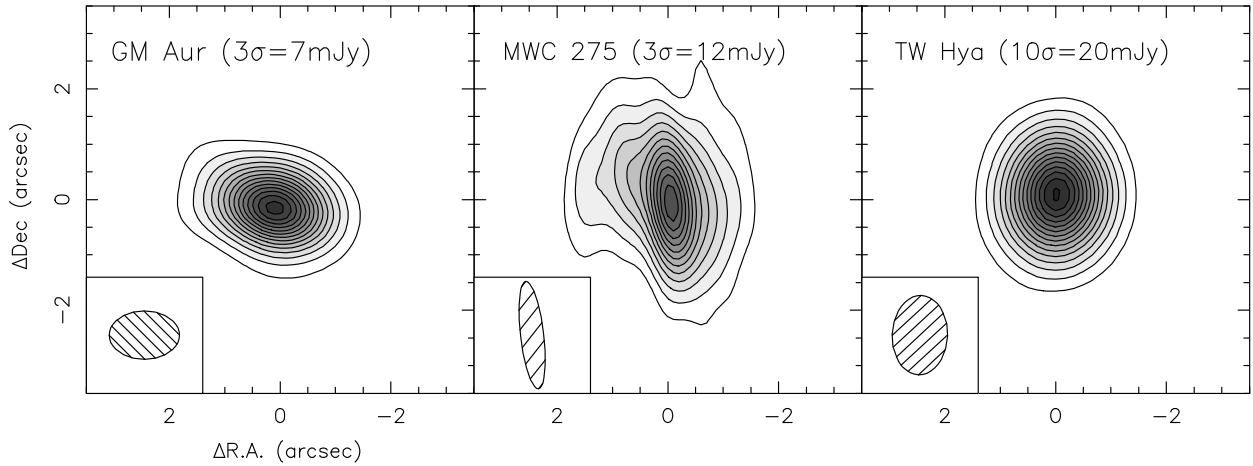}
\caption{\label{fig:cont_2} 1.3~mm dust continuum maps of GM~Aur, TW~Hya 
and MWC~275 \citep[from][]{hu08,is07}. For GM~Aur and MWC~275 contours begin at the 3$\sigma$ level and are
separated by the same amount. For TW~Hya contours start at and are separated by the 10$\sigma$ level.
Beam sizes and PA are listed in Table~\ref{tab:obs_1mm}. 
Integrated fluxes and source sizes are given in Table~\ref{tab:gau_fit}. }
\end{figure*}

Details of the 1.3 mm SMA dust continuum observations and data calibration 
of GM~Tau and TW~Hya can be found in \citet[][and references therein]{hu08}.
The angular resolution of these observations was about 1\arcsec\ at 1.3 mm. 
Similarly, details on the
 PdB observations of MWC~275 can be found in \cite{is07}. In this case the 
resolution was  2\arcsec$\times$0.4\arcsec. 
The 1.3~mm maps of these three sources are shown in Figure~\ref{fig:cont_2}.

\section{Observational results}
\label{sec:obs_res}
The observed continuum emission toward each of the disks is spatially resolved, 
centered on the position of the parent star and shows an almost centro-symmetric 
surface brightness profile. In Table~\ref{tab:gau_fit} we report the 
spatially integrated flux (column 2), the angular size obtained by fitting 
the continuum map with a bidimensional Gaussian profile (column 3), a first 
order derivation of the disk inclination measured from the aspect ratio 
of the Gaussian fitting (column 4), and the position angle measured East 
from North (column 5). Column 6 shows the disk radius, $R_g$, 
defined as the radius containing 95\% of the observed emission. In practice,
$R_g\sim 1.4 \times \textrm{FWHM}$ measured along the apparent disk major axis 
assuming the stellar distances discussed in \S~\ref{sec:sample}. 
Given the sensitivity of the observations and the fact that the dust emission
 is optically thin, this radius provides only a rough 
estimate of the real extent of the disk. 
 
\begin{figure*}
\centering
\includegraphics[angle=0,width=13cm]{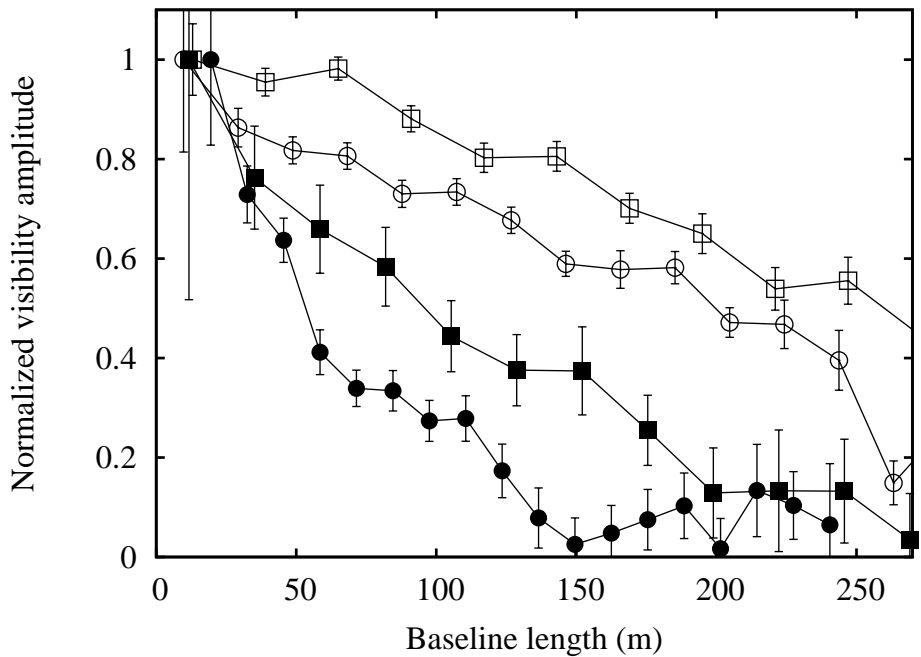}
\caption{\label{fig:amp} Normalized visibility amplitude as a function of 
baseline length for DR~Tau (open squares), RY~Tau (open circles), 
LkCa~15 (filled squares) and GSS~39 (filled circles).
}
\end{figure*}

The measured integrated fluxes are in good agreement (20\%) with earlier 
interferometric observations  at lower angular resolution, 
suggesting that only a small fraction of the flux is emitted on completely 
resolved out disk scales. We therefore assume that the CARMA observations 
trace the bulk of the disk emission. Disk radii range from 90 AU (DR~Tau) to 
320~AU (SR~24~S) and the surface brightness slopes vary from steep (e.g., DG~Tau, 
DR~Tau, DN~Tau) to quite shallow (e.g., RY~Tau, CY~Tau, LkCa~15, GSS~39). 
This latter point is illustrated in Figure~\ref{fig:amp}, which shows 
the normalized visibility amplitude $V$ as a function of the baseline length between
0 and 270~m for four disks. The most extended source, GSS~39 (filled 
circles), is completely resolved out on a baseline of 150~m, corresponding 
to an angular scale of about 1.8\arcsec, while the most compact, DR~Tau 
(empty squares), is resolved out only on a baseline of about 400 m, 
corresponding to an angular scale of $\sim$0.7\arcsec. RY~Tau (empty 
circles) and LkCa~15 (filled squares) are intermediate cases, 
resolved on angular scales of 0.8\arcsec\ and and 1\arcsec\ respectively. 

Given the high S/N ratio of the observations and the good image quality, 
the differences in the dust emission morphology may be reasonably 
attributed to different radial dust properties. Assuming that 
the dust emission is optically thin (as discussed in 
\S~\ref{sec:prof} this is a good approximation for most of our 
objects), the observed surface brightness is proportional to the emitted 
flux expressed by
\begin{equation}
\label{eq:fnu}
F_{\nu}(R) \propto \Sigma_d(R) \cdot k_{\nu}(R) \cdot  T(R),
\end{equation}
where $\Sigma_d$ is the dust surface density, $\nu$ is the frequency of the 
observation, $k_{\nu}$ is the dust opacity at the frequency $\nu$, $T$ is the 
dust temperature, and $R$ is the orbital radius. Once corrected for the 
stellar distance, the disk inclination and the synthesized beam shape, 
different surface brightness profiles can result from different radial 
profiles of the dust density, opacity and/or temperature.

On the following, we will compare the observed disk emission with 
theoretical models to (i) derive the dust properties, particularly density 
and temperature, from the available observations, (ii) quantify the differences 
between disks and (iii) investigate the origin of the different dust emission 
morphologies.


\section{Disk model and data analysis}
\label{sec:desc}
To analyze the observed disk emission we compare the measured complex 
visibilities with a theoretical model based on the {\it two layer} 
approximation \citep{cg97} and on the similarity solution for the disk 
surface density of a thin 
keplerian viscous disk \citep{lb74,pr81,har98}. The basic properties of the 
model are summarized in \S~\ref{sec:mod}.

The disk model that best fits the observations is found by adopting 
$\chi^2$ as the maximum likelihood estimator, where $\chi^2$ is defined by
\begin{equation}
\label{eq:chi2}
\chi^2 = \sum \left[ (Re_o^2 - Re_t^2)+(Im_o^2 -Im_t^2)\right]\cdot w;
\end{equation}
$Re$ and $Im$ are the real and imaginary part of the observed (suffix 
$_o$) and theoretical (suffix $_t$) complex visibilities. The weight $w$ 
of each data point in the {\it uv} plane is given by
\begin{equation}
\frac{1}{\sqrt{w}}=\sigma = \frac{2k_b T_{sys}}{\eta_s \eta_a A 
\sqrt{2 \Delta\nu \tau_{acc}}}
\end{equation}
where $k_b$ is Boltzmann's constant, $T_{sys}$ is the system 
temperature, $A$ is the antenna area, $\Delta\nu$ is the band width, 
$\tau_{acc}$ is the integration time on source and $\eta_{s}$ and $\eta_{a}$ 
are the system and antenna efficiencies respectively. To minimize $\chi^2$ and 
evaluate the constraints on the model parameters we use a Markov Chain 
Monte Carlo method as described in Appendix~\ref{sec:MCMC}.

\subsection{Disk surface density} 
\label{sec:modsim}

The most common approach to derive the disk surface density distribution 
from millimeter and sub-millimeter observations has been to adopt a power 
law parameterization of $\Sigma$ in the form
\begin{equation}
\label{eq:sigma_power}
{\Sigma}(R)= \Sigma_1 \left( \frac{R_1}{R} \right)^p,  
	\;\;\; \textrm{with}\;\;\; R_{in}<R<R_{out},
\end{equation}
where $\Sigma_1$ is the surface density value at an arbitrary radius 
$R_1$, and $R_{in}$ and $R_{out}$ are the disk inner and outer radii. 
This parameterization was initially motivated by the 
empirical results of \cite{hay81}, who derived 
a power law surface density distribution for the solar nebula with 
$p=1.5$ between 0.35 and 36 AU. This surface density parameterization 
has been recently revised to accommodate a number of theoretical and 
observational issues. First, there is no physical 
justification for a power law disk surface density in terms of disk 
formation and evolution \citep[see, e.g.,][]{hg05}. Moreover, 
such a distribution must be artificially limited between an 
inner and outer disk radius to obtain a finite disk mass.
In addition, it has been demonstrated that the power law 
parameterization fails to explain the differences in the 
radial extensions of the dust and the gas emission 
that are observed in a number of intermediate mass pre-main sequence stars 
\citep[][hereafter H08]{pi05,is07,hu08}. Independently it has been suggested
that the gas distribution in the early solar system would be better 
explained by a surface density distribution of the form 
$\Sigma(R) \propto R^{-1/2} \times \exp(-R^{3/2})$ \citep{d05}.

Following H08, we adopt the similarity solution of the surface density
of a thin keplerian disk subject to the gravity of a point mass $M_{\star}$
\citep{pr81,lb74} in the form presented by \cite{har98}:
\begin{equation}
\label{eq:sigma}
\Sigma(r,t)= \frac{C}{3\pi \nu_1 r^\gamma}\tilde{t}^{-\frac{(5/2-\gamma)}
	{(2-\gamma)}}\exp{\left[-\frac{r^{(2-\gamma)}}{\tilde{t}} \right]}
\end{equation} 
where $C$ is a normalization constant, $r$ is the stellocentric distance 
expressed in the units of a radial scale factor $R_1$ ($r=R/R_1$), 
$\nu_1$ is the disk viscosity at radius $R_1$, $\gamma$ is the slope of the 
disk viscosity $\nu(R) \propto R^\gamma$, $\tilde{t}$ is the non-dimensional 
time, $\tilde{t}=t/t_s+1$, $t$ is the age of the disk 
and $t_s$ is the disk viscous time at the radius $R_1$ defined by
\begin{equation}
\label{eq:ts}
t_s=\frac{1}{3(2-\gamma)^2} \frac{R_1^2}{\nu_1}.
\end{equation}
Based on Eq.~\ref{eq:sigma} we demonstrate in 
Appendix ~\ref{sec:mass} that $R_1$ is the radius containing 63\% of the 
disk initial mass (at $t=0$). 

As pointed out by H08, this form of the 
surface density has the particular characteristic of falling off 
exponentially at large disk radii, thereby providing sufficiently 
dense gas in the outermost disk regions to explain the observed radial 
extent of the gas emission. Moreover, since the similarity solution relates the 
surface density in Eq.~\ref{eq:sigma} to the age of the system, we can now 
investigate the details of disk evolution using our 
millimeter wave observations of the disk emission.

The surface density as expressed by Eq.~\ref{eq:sigma} includes a significant 
number of unknown quantities, $C$, $\nu_1$, $\gamma$, $t_s$ and $R_1$, 
which cannot be constrained by the observations. A form more suitable 
for model fitting can be written by taking the derivative of Eq.~\ref{eq:sigma} 
with respect to time and introducing the mass flow 
\begin{multline}
\label{eq:mass_flow}
\Dot{M}(r,t)=C \tilde{t}^{-\frac{(5/2-\gamma)}{(2-\gamma)}} \exp{\left
[-\frac{r^{(2-\gamma)}}{\tilde{t}} \right]} \\ \times \left[ 1- \frac
{2(2-\gamma)r^{(2-\gamma)}}{\tilde{t}} \right]
\end{multline}
Since disk evolution is governed by the conservation of the 
total angular momentum, the disk must expand while matter is accreting 
on the central star. Thus $\Dot{M}$ must change sign at a
{\it transition radius}, $R_t$, where
\begin{equation}
\label{eq:Rt}
R_t \equiv R_1 \left[\frac{\tilde{t}}{2(2-\gamma)} \right]^{1/(2-\gamma)}.  
\end{equation}
The resulting mass flow is directed inward for $R<R_t$ (accretion) and outward (expansion)
for $R>R_t$. 

\begin{figure*}
\begin{center}
	\includegraphics[angle=0,width=12cm]{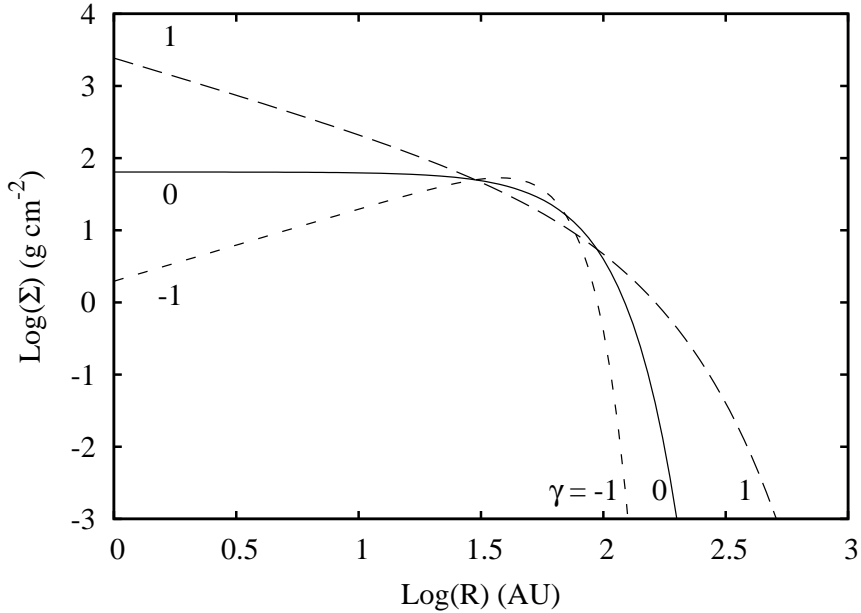}
	\caption{\label{fig:sigma} Disk surface density $\Sigma(R)$ 
	as defined by Eq.~\ref{eq:sigma_used} for $\gamma$ = -1, 0 and 1. 
	The transition radius and the normalization 
	are fixed at $R_t = 30$ and $\Sigma_t = 50$ g/cm$^2$.}
\end{center}
\end{figure*}

The surface density $\Sigma$ can be rewritten in the form
\begin{multline}
\label{eq:sigma_used}
\Sigma(R,t) = \Sigma_t \left( \frac{R_t}{R} \right)^{\gamma} \\ \times  \exp{ \left\{ - \frac{1}{2(2-\gamma)} 
\left[ \left( \frac{R}{R_t} \right)^{(2-\gamma)} -1 \right] \right\} }
\end{multline}
where we adopt the physical radius $R$ and group all the other 
unknown quantities within $\Sigma_t$ (i.e., the surface density at the 
radius $R_t$). Figure~\ref{fig:sigma} displays the behavior of the 
surface density for different values of $\gamma$. When $\gamma=0$, 
Eq.~\ref{eq:sigma_used} becomes a Gaussian law, while for 
negative $\gamma$ the surface density has a maximum at 
\begin{equation}
\label{eq:sigmax}
R_{max} = R_t \times (-2\gamma)^{1/(2-\gamma)}.
\end{equation}
Our new spatially resolved observations enable us to constrain 
$\Sigma_t$, $R_t$ and $\gamma$ and define the disk surface density 
distribution.

\subsection{Disk structure and emission}
\label{sec:mod}

The flux emitted by the circumstellar dust can be computed by solving 
for the structure of a passive keplerian disk (i.e., one that is heated 
only by the stellar radiation) adopting the {\it two-layer} approximation 
of \citet{cg97}. If the disk is vertically optically thick to the stellar 
radiation, its thermal structure is characterized by a surface layer 
temperature $T_s$, which is appropriate for regions where the optical 
depth to the stellar radiation is $<$ 1, and by a disk interior temperature 
$T_i$, characteristic of deeper disk regions. Both 
temperatures can be calculated as function of the orbital radius $R$ 
by iterating on the vertical disk structure \citep[see][]{d01}. 
Assuming hydrostatic equilibrium between the 
gas pressure and the stellar gravity, the disk has a flared geometry 
with the opening angle increasing with the distance from the star and the 
vertical gas distribution expressed by a Gaussian law normalized to the 
surface density distribution described by Eq.~\ref{eq:sigma_used}.  

Due to the radial exponential fall-off of the disk surface 
density, the very outermost disk regions are optically thin to the stellar 
radiation and the {\it two-layer} approximation can not be applied. For 
densities typical of TTS disks, the transition to this optically thin regime
occurs at a radius $R_d$ which is much larger than the transition radius $R_t$. 
Given the dust density and temperature for $R>R_d$, we adopt 
$R_d$ as the disk outer radius with negligible effects on the strength of the 
overall dust emission.

Once the disk thermal structure is known, the continuum dust emission can be  
computed by combining the flux arising from the disk interior $F_\nu^i$ with the flux 
from the disk surface layer $F_\nu^s$. These are expressed respectively by 
\begin{multline}
F_\nu^i = 2\pi \cos{i} \int_{R_{in}}^{R_d} \left\{1-\exp \left[ \frac{-\Sigma(R) 
k^i_\nu}{\cos{i}} \right] \right\} \\
\times B_\nu[T_i(R)] \, \frac{R}{d^2} \, dR
\end{multline}
and 
\begin{multline}
\label{eq:fs}
F_\nu^s = 2\pi \int_{R_{in}}^{R_d} \left\{1+\exp \left[ \frac{-\Sigma(R) 
k^i_\nu}{\cos{i}} \right] \right\} \\
\times B_\nu[T_s(R)] \, \Delta\Sigma(R) \, k^s_\nu \, \frac{R}{d^2} \, dR
\end{multline}
\citep{d01,c01}, where $\nu$ is the frequency, $d$ is the distance to 
the source, $i$ is the disk inclination with respect to the plane of 
the sky ($i=0$ for face-on), $B_\nu(T)$ is the Planck function, $\Delta\Sigma$ is 
the column density in the disk surface and $k_\nu^{i,s}$ are the dust 
opacities at the disk mid plane and surface, as discussed in the next 
section. The disk inner radius $R_{in}$ is fixed 
at the dust evaporation distance and varies between $\sim$0.03 and $\sim$0.5 AU 
for the stellar luminosities characteristic of our sample \citep{in05, is06}.

\subsection{Dust opacity}
\label{sec:dust_opa}
\begin{figure*} 
\centering
\includegraphics[angle=0,width=12cm]{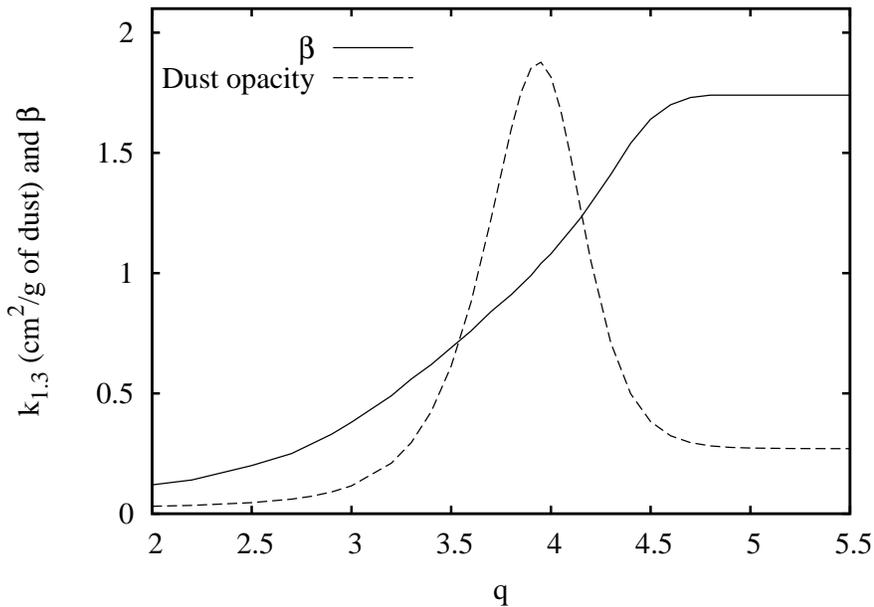}
\caption{\label{fig:opac} Dust opacity at 1.3~mm ($k_{1.3}$) and the slope $\beta$ 
of the dust opacity at millimeter wavelengths as a function of the slope, $q$, 
of the grain size distribution $n(a) \propto a^{-q}$. A minimum grain 
radius of 0.01~$\mu$m and a maximum grain radius of 10~cm are assumed.}
\end{figure*}

To calculate the disk structure and emission we adopt the optical constants 
of astronomical silicates and carbonaceous materials \citep{wd01,zu96}. 
The dust opacity is calculated assuming compact spherical grains and 
adopting the fractional abundances used by \citet{p04} and results in 
a dust/gas ratio close to 0.01. We assume a grain size distribution of 
the form $n(a)\propto a^{-q}$ between 0.01~$\mu$m and 10 cm, where $q$ is 
a free parameter of the model. 
Figure~\ref{fig:opac} shows the dust opacity at 1.3 mm ($k_{1.3}$, dashed 
line) and the slope $\beta$ of the dust opacity ($k_\lambda \propto \lambda^{-\beta}$)  
calculated between 1 
and 7~mm (solid line), as a function of the slope of the grain size 
distribution $q$.
$k_{1.3}$ reaches a maximum value of 1.9 cm$^{2}$ {\it per gram of dust} for 
$q=3.9$ and decreases to values smaller than 0.4~cm$^{2}$ {\it per gram of dust} 
for $q<3.4$ and $q>4.5$. The slope $\beta$ increases with $q$, and varies 
between $\beta=0.1$ for $q=2$ (when the opacity is dominated by 10 cm size grains) 
to $\beta=1.7$ for $q>4.5$ (when the opacity is dominated by sub-micron grains; 
see Natta et al.~2007 and references therein for more details on the variation of 
$\beta$ with the grain size and composition). 

To solve for the disk structure, we adopt different values of $q$  for the disk 
interior and the disk surface layer. To a first approximation, as long as the disk 
is optically thick to the stellar radiation neither the disk structure or 
the millimeter-wave dust emission depend on the assumed dust opacity in the 
disk surface layer. We therefore fix the value of $q$ at 5 in the disk surface so that the 
opacity is dominated by the sub-microns grains which are generally required 
to explain the silicate features observed between 10 and 20 $\mu$m \citep{fur06}. 
The value of $q$ in the disk interior is a free parameter and is chosen so as
to reproduce the measured slope $\alpha$ of the spectral energy distribution 
($F_\nu \propto \nu^{\alpha}$) between 0.8 and 7 mm. The resulting values of 
$\alpha$, $q$, $k_{1.3}$ and $\beta$ are presented and discussed in \S~\ref{sec:dust}.

It is important to emphasize that since we are analyzing the spatially
resolved dust emission observed at a single wavelength, we cannot disentangle the radial 
variation of the dust opacity from the variation of dust surface density. 
In effect, the observations constrain the product $k_{1.3}(R) \times \Sigma(R)$ 
(see \S~\ref{sec:obs_res}). Since the radial
variation of the dust opacity has not been quantified observationally, 
we assume that $k_{1.3}$ is constant throughout the disk.

\section{Results of the model fitting}
\label{sec:res}
\begin{figure*}
\begin{center}
	\includegraphics[angle=0,width=\textwidth]{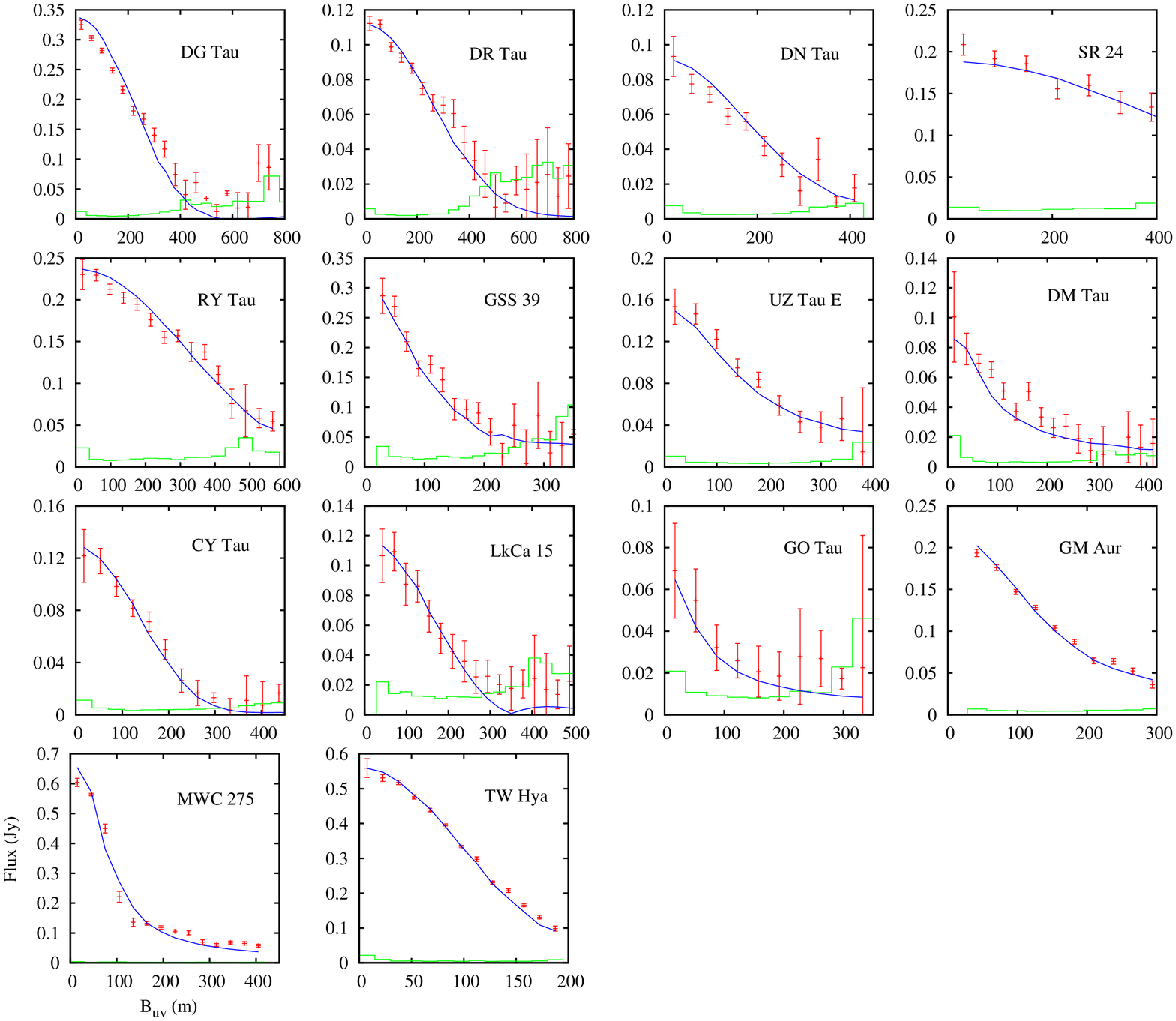}
	\caption{\label{fig:vismodel} Comparison between the observed correlated flux (dots) 
          and the best fit model prediction (solid line) as a function of the beseline lenght
          deprojected using disk inclinations and position angles listed in Tab.~\ref{tab:res}.
          The histogram in the lower part of each panel shows the expected signal in case of
          zero flux.}
\end{center}
\end{figure*}

The model comparison with observations is described in Appendix~\ref{sec:MCMC}.
For all the objects we obtain good fits to the observations, with reduced 
$\chi^2$ close to 1. The dust properties adopted in the model fitting 
are presented in the Table~\ref{tab:dust}, which list the slope $q$ of the
grain size distribution (column 4) and the corresponding dust opacity at 1.3 mm
$k_{1.3}$ (column 5). Disk inclination, the position angle of the projected 
disk major axis, the transition radius $R_t$, the surface density $\Sigma_t$ 
at $R_t$ and the value of $\gamma$ corresponding to the best fit models are
shown in Table~\ref{tab:res}. The last two columns of Table~\ref{tab:res}
show the total disk mass and the radius $R_d$ at which 
the disk becomes optically thin to the stellar radiation.
For each object, the radial profiles of the disk surface density $\Sigma(R)$, 
the cumulative disk mass $M_d(R)$, the disk surface and interior 
temperature profiles $T_{s,i}(R)$ and the cumulative emission at 1.3~mm 
$F_{1.3}(R)$ are shown in Figure~\ref{fig:disk}. The comparison between
the observed and the best fit model visibility profiles are shown in 
Figure~\ref{fig:vismodel}, and the residuals, defined as the difference 
between the observed maps of the 1.3~mm dust emission and the 
models, are shown in Figure~\ref{fig:res_cont}.

\subsection{Properties of the dust}
\label{sec:dust}
\begin{figure*}
\centering
\includegraphics[angle=0,width=\textwidth]{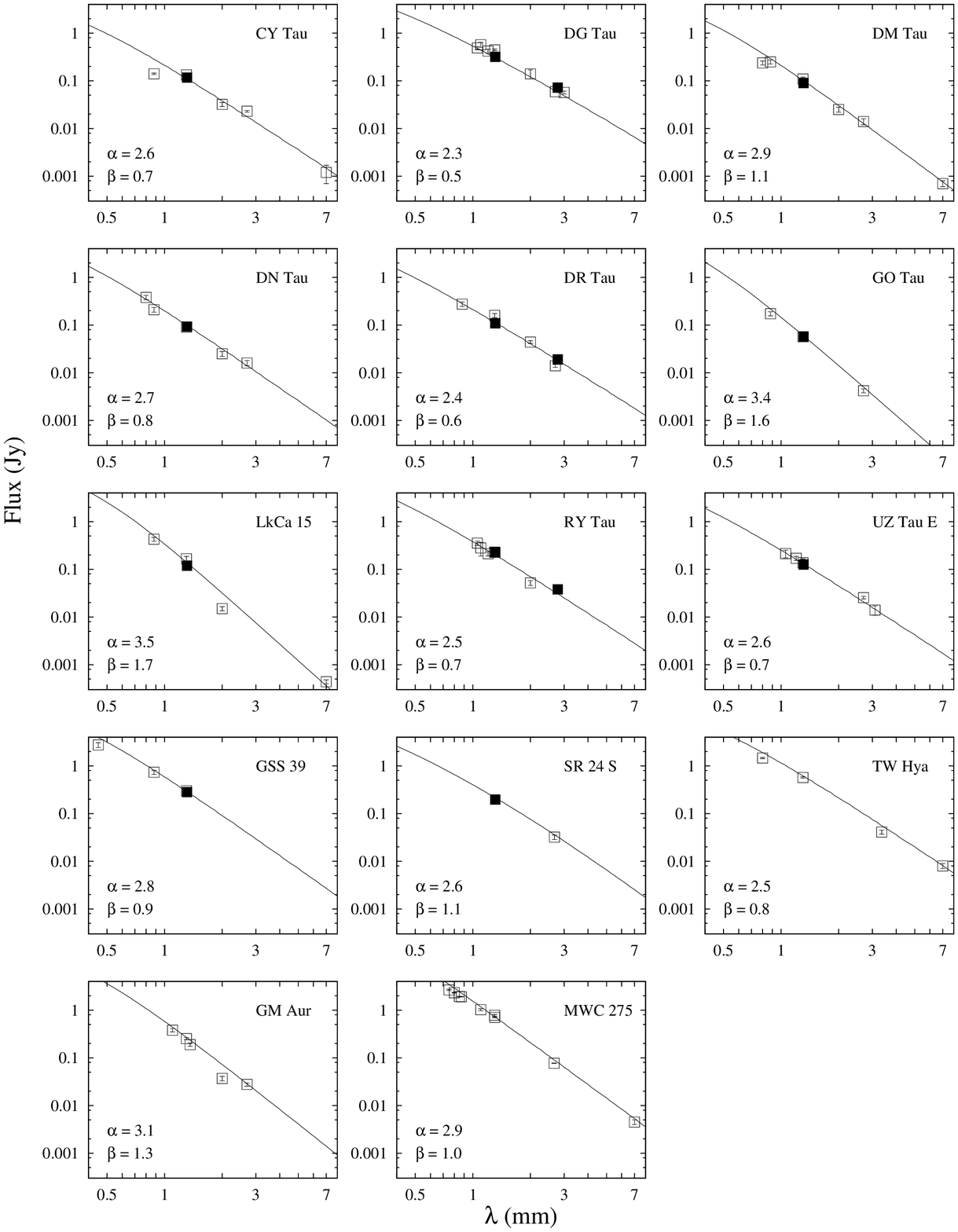}
\caption{\label{fig:slope} Flux density distribution of the observed sample 
between 0.4 and 8 mm. Filled squares correspond to the CARMA observations 
discussed in this paper, and open squares correspond to measurements available 
in literature \citep{d96,kit02,and05,rod06,is07}. The solid curves correspond 
to disk models that best fit the spatially resolved observations at 1.3~mm 
(see Table~\ref{tab:res}). For each object the values of $\alpha$ and of the 
dust opacity slope $\beta$ are indicated in the relative panel.}
\end{figure*}

Figure~\ref{fig:slope} shows the spatially integrated fluxes 
of the observed disks between 0.5 and 7 mm. CARMA observations 
are represented by filled squares while open squares depict 
data from the literature. The solid curves show the best 
fit models characterized by the values of spectral index 
$\alpha$ ($F_\nu \propto \nu^\alpha$), slope $\beta$ of the dust 
opacity, slope $q$ of the grain size distribution and dust 
opacity at 1.3 mm $k_{1.3}$ reported in column 2-5 of Table~\ref{tab:dust}.
We derive values of $\alpha$ between 2.4 (DG~Tau) to 3.5 (LkCa~15), 
which lead to values of $\beta$ between 0.5 and 1.7. If 
the disk emission is optically thin at 1.3 mm, $\alpha$ and 
$\beta$ are tightly correlated with $\beta=\alpha-2$ \citep[e.g.,][]{bs91}.
In practice, the denser regions of the disk are partially
optically thick at 1.3~mm and  $\beta \geq \alpha-2$. The difference 
between $\beta$ and $\alpha-2$ depends on the ratio of optically 
thick to optically thin emission from the disk and is only few 
percent in CY~Tau, DN~Tau, UZ~Tau~E and GSS39. However, it 
increases to about 20\% in SR~24~S and DG~Tau, where the disks 
are optically thick to the 1.3 mm emission within a radius of 
17~AU and 25~AU respectively. 

The dust opacity $k_{1.3}$ obtained by the SED fitting 
strongly depends on the assumed grain size distribution discussed
in  \S~\ref{sec:dust_opa}. Actually, we can reproduce the observed 
spectral indexes $\alpha$ with very different dust opacities if we fix
the slope of the grain size distribution $q$ and keep $a_{max}$ as a
free parameter. Values of $a_{max}$ and $k_{1.3}$ for the case 
$q=3$ are shown in the last two columns of Table~\ref{tab:dust}. 
This latter  grain size distribution leads to dust opacities larger than
the case with fixed $a_{max}$ and variable $q$ by  a factor 2-20.
As a consequence, much smaller surface densities 
and disk masses are required to reproduce the observed dust emission. 

Independently on the assumed grain size distribution, values of $\beta \lesssim 1$ 
imply that the dust opacity is strongly influenced by dust grains larger 
than 1~mm \citep[e.g.][]{nat07}. This suggests that the circumstellar 
dust around CY~Tau, DG~Tau, DN~Tau, DR~Tau, RY~Tau, UZ~Tau~E, GSS~39 
and TW~Hya has undergone important grain growth processes. However, the 
small value of the spectral index for DG~Tau can now be probably 
explained by the strong contribution of the optically thick emission 
at 1.3~mm. By contrast the dust properties in LkCa~15 and GO~Tau 
are more similar to that found in the ISM. 
Moreover, disks with small value of $\beta$ tend to be characterized by
small value of $\gamma$ and therefore to steeper surface density profiles 
for $R>R_t$. This suggest that the dust surface density may be correlated with 
the dust properties and in particular that small and large dust grains might
have different radial density distributions as suggested by recent theoretical 
models of growth and radial migration of dust grains in proto-planetary 
disks \citep{br08}.


\subsection{Radial profile of the surface density}
\label{sec:prof}
In the panels of the first and second columns of 
Figure~\ref{fig:disk} we show for each object the surface density and 
the cumulative flux at 1.3 mm (solid line), the 1-$\sigma$ uncertainty 
range (shaded region) and the spatial resolution provided by our interferometric 
observations (vertical dashed line). The dust emission coming from 
spatially resolved disk regions varies from about 85\% (CY~Tau, LkCa~15, 
MWC~275, GSS~39) to less than 30\% (DR~Tau, SR~24~S) of the total flux. 
As a consequence, the constraints on $R_t$, $\gamma$ and $\Sigma_t$ vary 
from few percents to about 30-40\%. 

The derived transition radii range from 17.5 and 110 AU. For the 
assumed dust opacities and stellar distances, the radius $R_d$ 
at which the disk becomes optically thin to the stellar radiation - in 
effect the disk outer radius (\S~\ref{sec:mod}) - varies from 73~AU in 
the case of TW~Hya to 670~AU for GO~Tau. These values are close to the 
disk outer radii inferred from the analysis of the optically thick CO 
emission \citep{sim00}, but larger than existing determinations of the 
disk outer radii based on power law surface density models \citep{and07}.
In the case of MWC~275, we find $R_d = 520$ AU. This radius is 
in very good agreement with the gas extent inferred from the 
CO line emission but it is twice as large as the disk outer radius 
implied by the same observations if a power law surface 
density disk model is adopted \citep{is07}. It appears that there is 
in fact {\it no} discrepancy between the radial extents of the gas and 
dust emission. The outer radii determination of the dust extent based 
on the exponential fall-off for the surface density are quite comparable 
with the radii derived from observations of the CO emission (see also H08). 

From Eq.~\ref{eq:sigma_used}, it is clear that the disk surface density
is characterized by $\gamma$, which ranges from -0.8 to 0.8. 
For $R \gtrsim R_t$, radial density profiles become steeper with decreasing 
$\gamma$, as shown in Figure~\ref{fig:sigma}. This is further illustrated by 
the plots of $\Sigma$ in the panels of the first column of Figure \ref{fig:disk}. 
In Figure~\ref{fig:disk_a}, LkCa~15, CY~Tau, DG~Tau and TW~Hya, have negative 
values of $\gamma$ with a probability of at least 68\% (1-$\sigma$, see 
Appendix~\ref{sec:MCMC}). From Eq.~\ref{eq:sigmax} (\S~\ref{sec:modsim}), 
the corresponding surface densities have maxima at 89, 41, 21 and 13~AU 
respectively. For LkCa~15, the CARMA observations clearly resolve the surface 
density maximum and the almost flat profile of $\Sigma$ between 50 and 100 AU. 
For the other objects the predicted maxima lie inside the resolution of our 
observations. Disks with values of $\gamma \sim 0$ are shown in 
Figure~\ref{fig:disk_b}, and for $\gamma > 0$ in \ref{fig:disk_c}. For almost 
all disks, $\Sigma$ is well constrained for $R \gtrsim 40$ AU. For DR~Tau and 
SR~24~S, almost 80\% of the observed emission arises from the innermost 
spatially unresolved disk region so that the disk structure is poorly 
constrained. The structure of GO Tau disk is also poorly constrained due to 
the low S/N ratio of the observations.

The inferred disk structure depends weakly on the adopted grain size distribution 
if the SED is used to constrain the dust opacity. 
In fact, the disk mid plane temperature $T_i(R)$ varies by less than 5\%
between the two different grain size distribution models shown in Table~\ref{tab:dust}. 
Consequently, since $\Sigma(R) \propto T^{-1}(R)$, 5\% is also the maximum 
variation observed in the profile of the surface density. We obtained model fits for a
subset of the disks using the alternative grain size distribution listed in 
Table~\ref{tab:dust}, and have verified that the variations on 
$\gamma$ and $R_t$ are much smaller that the respective uncertainties.

\subsection{Disk mass}
\label{sec:diskmass}
The radial integration of the disk surface density leads to the cumulative disk 
mass $M(R)$ presented in the panels in column 3 of Figure~\ref{fig:disk}, 
and to the total disk mass $M_d$ in Table~\ref{tab:res}. 
For the adopted grain size distribution  and dust/gas ratio 
(\S~\ref{sec:dust_opa}), $M_d$ varies by more than a order of magnitude 
from $\sim$0.02 to $\sim$0.4 $M_\sun$, with uncertainties between 
20\% and 200\% depending on how well $\Sigma$ is constrained.
$M_d$ is. In the case of DG~Tau, the mass of the disk is comparable
with the stellar mass. Note however that stellar masses derived using DM97
stellar evolution models are probably a lower estimate of the real stellar
mass. Indeed, the disk mass is only 35\% of the
stellar mass if \cite{bar98} models are used to derive the stellar mass from 
the HR diagram (see Appendix~\ref{sec:HR}). Disk mass is 15\% of the mass of 
the parent star for DN~Tau and GM~Aur, while it is only a few percent for all 
the other objects.

Disk masses strongly depend on the dust opacity at 1.3 mm which may vary by 
a large factor if a different grain size distribution is adopted. The case $q=3.0$ 
discussed is Section~\ref{sec:dust} leads to disk masses between a factor 2 and 20
smaller than what discussed above. This introduces an additional large 
uncertainty on the disk masses derived from millimeter-wave observations. 

As shown in the cumulative mass plots, most of the disk mass is concentrated 
in the outermost disk regions, independent of the exact surface density 
profile. By contrast, since the disk scale height $H(R)$ increases with 
distance from the star, the dust+gas volume density in the disk mid-plane 
[$\rho_0 \propto \Sigma(R)/H(R)$] is at a maximum close to the inner disk radius. 
A comparison of $M_d$ with the minimum mass of the solar nebula
(0.02 $M_{\sun}$, see Figure \ref{fig:disk}) shows that this amount of material 
is contained within radii ranging from $\sim$10 AU (for DG~Tau) to $\sim$60 AU 
(for DM~Tau, MWC~275). These values are not too far from the minimum mass 
solar nebula outer radius of 36~AU postulated by Hayashi (1981). We will return to 
this point in \S~\ref{sec:solar}.

\subsection{Disk temperature}
\label{sec:disktemp}
The panels in the last column of Figure~\ref{fig:disk} display the 
radial profile of the dust temperature in
 the disk interior, $T_i$ (thick line), and in disk surface layer, $T_s$ 
(thin line). As discussed in \S~\ref{sec:mod}, $T_s$ scales roughly 
as $R^{-1/2}$ due to the dilution of the stellar radiation and to 
the variation of the Planck mean dust opacity. It varies between $\sim$1500 K 
at the disk inner radius, and 10 to 40 K at $R_d$, where the disk becomes 
optically thin to the stellar radiation in the vertical direction. Since 
the dust column density in the disk atmosphere is a tiny fraction 
($\sim10^{-4}$) of the total disk surface density, the contribution to 
the observed  millimeter emission from the disk atmosphere is 
negligible (see Eq.~\ref{eq:fs}).

Inside a radius of 50-100 AU,  the dust temperature $T_i$ in the deeper disk 
region also scales as $R^{-1/2}$, as expected for an irradiated 
optically thick disk \citep{kh95}. At larger distances the dust density is
low enough that the disk interior becomes progressively 
optically thin to the emission of the disk atmosphere (which heats up the disk
interior) and to its own thermal emission (which cools the disk interior). 
The resulting effect is that the temperature profile deviates from 
the $R^{-1/2}$ relation. The deviation is larger in disks characterized by
small values of $\gamma$ and a correspondly steeper decrease in the
surface density (see Figure~\ref{fig:disk_a} and \ref{fig:disk_b}). In these 
objects, the temperature of the disk region that dominates the 
observed millimeter emission assumes almost constant values between 20 and 
40 K. By contrast, $T_i$ decreases monotonically in disks with the
highest $\gamma$ (i.e., DM~Tau, GO~Tau and UZ~Tau~E).  In no cases does 
the disk temperature fall below 10 K which is generally assumed to be the
equilibrium temperature with the interstellar radiation field.

\subsection{Disk orientation}
\label{sec:disk_ori}
Disk inclinations and the position angles derived from the observations are 
uncorrelated and randomly distributed as expected in absence of a preferential 
disk orientation in space. For CY~Tau, DG~Tau, DN~Tau, DM~Tau, LkCa15, 
UZ~Tau~E, GM~Aur, GSS39, SR24, TW~ Hya and MWC 275 both inclinations and position 
angles are in agreement with published values within 2$\sigma$ 
\citep{j96,d97,sim00,qi04,pi06,and07,is07}. For GO~Tau, DR~Tau and RY~Tau we
derive disk inclinations of 25\degr$\pm$25\degr, 37\degr$\pm$3\degr and 
60\degr$\pm$3\degr respectively, considerably lower than the 
66\degr, 72\degr\ and 86\degr\ suggested by \citet{and07}. On the other hand 
our RY~Tau results agree well with recent optical 
observations \citep{sb08} suggesting that the discrepancies may be 
due to the lower S/N ratio of \cite{and07} observations compared with the 
new CARMA results.

\section{Discussion}
\label{sec:disc}

From the results of \S~\ref{sec:res}, we conclude that the diversity in the 
surface brightness profiles observed in our 1.3~mm dust continuum maps
(\S~\ref{sec:obs_res}) is due mainly to different surface 
densities, $\Sigma(R)$, with the disk temperature $T_i(R)$ having only a 
minor effect. Of course, this assumes that the dust opacity is constant 
with radius as discussed in \S~\ref{sec:dust_opa}. 
Most of the discussion below is therefore devoted to
examining the dust density profiles and investigating the 
possible sources of the observed variations.

\subsection{Disk evolution}
\label{sec:evol}

\begin{figure*}
\centering
\includegraphics[angle=0,width=11.5cm]{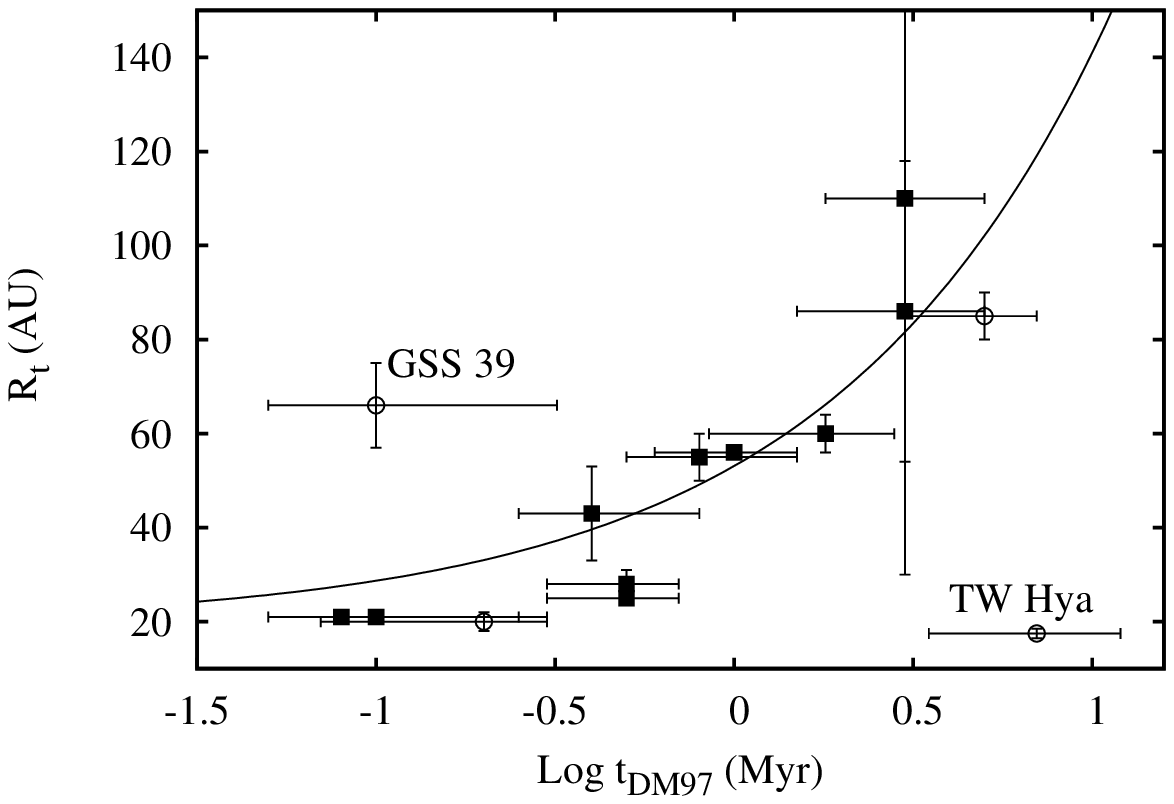}
\caption{\label{fig:Rt_tstar} The disk transition radius 
$R_t$ as a function of the stellar age computed using DM97 models.
Filled squares show the 10 disks located in Taurus-Auriga star forming region.
For this sub sample, the correlation coefficient between $R_t$ and the stellar
age is 0.98 and the probability that the data are randomly distributed is less than
0.1\%. The solid line corresponds to $R_t =  R_0 + C\cdot t^\eta$ 
with $\eta=0.5\pm 0.4$, $R_0 = 17\pm10$~AU and $C=37\pm20$.}
\end{figure*}

\begin{figure*}[]
\centering
	\includegraphics[angle=0,width=11.5cm]{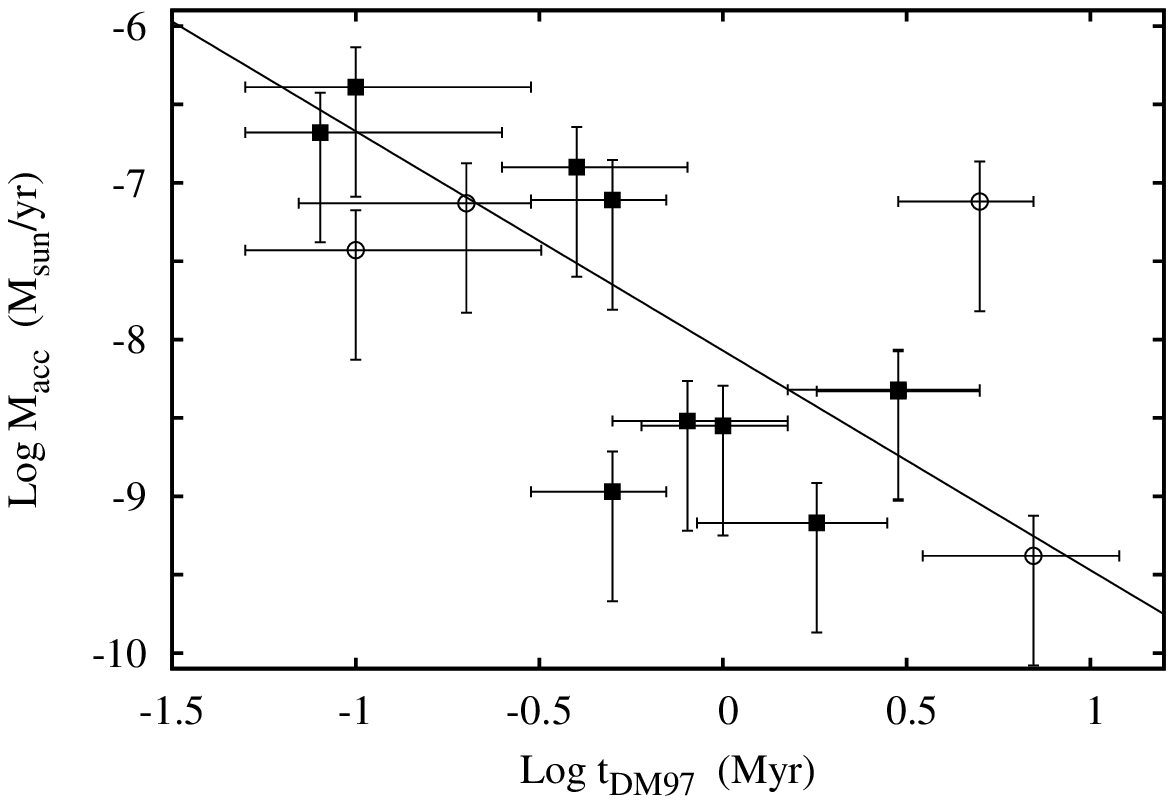}
	\caption{\label{fig:Macc_t} The mass accretion rate as a function of 
	stellar age computed from DM97 models. Filled squares show the 10 
	disks located in Taurus-Auriga star forming region. The correlation 
	coefficient between $\Dot{M}_{acc}$ and the stellar age is -0.62 and the
	probability that the data are randomly distributed is about 2\%. 
	The solid line corresponds to $\Dot{M}_{acc} \propto t^{-1.4}$.}
\end{figure*}

\begin{figure*}
\centering
\includegraphics[width=12cm]{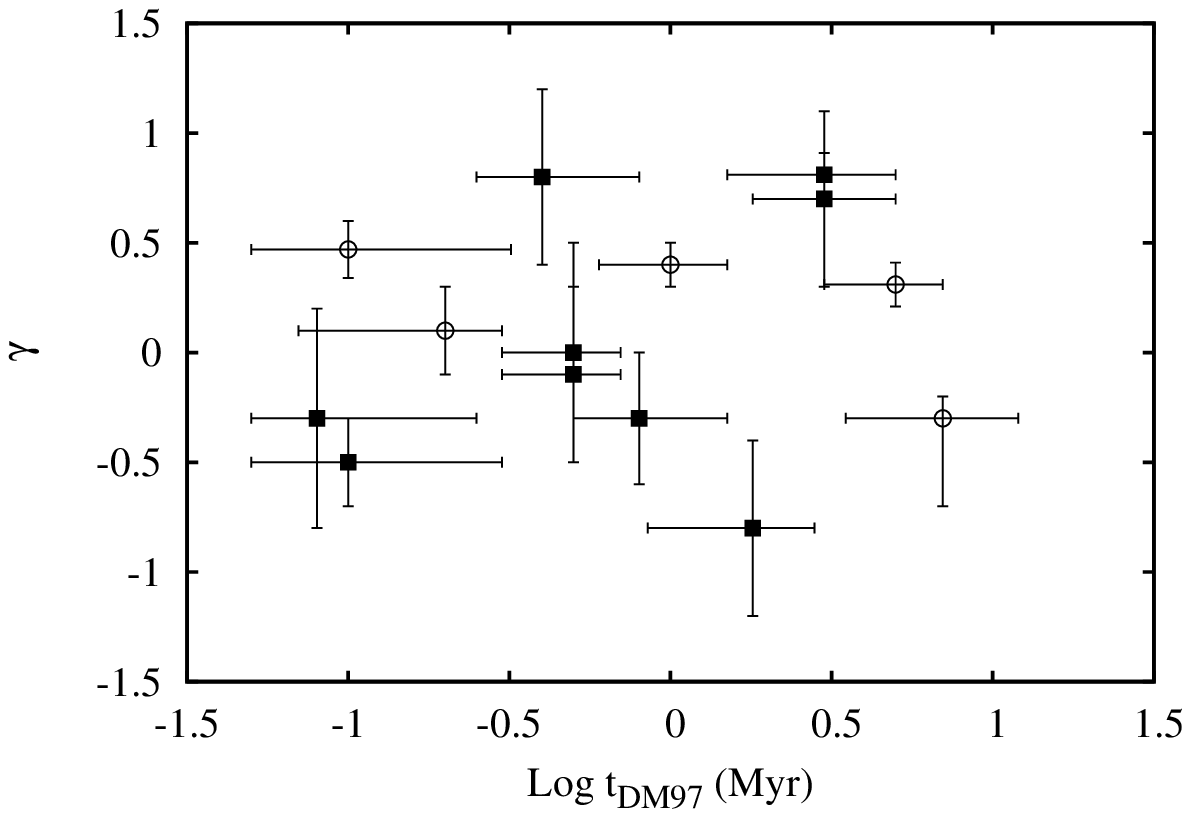}
\caption{\label{fig:gamma_tstar} Variation of the parameter $\gamma$, 
which defines the surface density profile, with the stellar age calculated
from DM97 models. The 10 disks in Taurus-Auriga are shown by filled squares.}
\end{figure*}

Figure~\ref{fig:Rt_tstar} compares the disk transition radius $R_t$ and 
the stellar age (from Appendix \ref{sec:HR}).
Over the $\sim$5 Myr span of our sample ages, $R_t$ appears to increase 
with stellar age from about 20 to 100 AU. Applying the 
non-parametric Spearman rank-order test (see, e.g., Press et al.~2007), 
the correlation coefficient is $r=0.42$ and the probability that 
the data are randomly distributed is 12\%. However, limiting the analysis 
to the objects in Taurus-Auriga (filled squares) the correlation coefficient 
increases to 0.98 and the probability of a random distribution falls below 0.1\%.
If $R_t$ has a power law dependence of the form $R_t = R_0 + C\cdot t^\eta$, 
where $t$ is the stellar age in Myr, we find that $R_t$ increases
as $\sqrt{t}$ , with $\eta=0.5\pm 0.4$, $R_0 = 17\pm10$~AU 
and $C=37\pm20$. As shown in Figure~\ref{fig:Macc_t},
this disk expansion is accompanied by a decrease in the mass accretion rate
roughly described by $\Dot{M}_{acc} \propto t^{-1.4\pm0.3}$.   
In this case, the Spearman test indicates that the probability that the 
data are randomly distributed is 2\% (r=-0.62) both for the full sample and 
for the objects in Taurus-Auriga. As illustrated in Fig.~\ref{fig:gamma_tstar},
there appears to be no correlation between $\gamma$ and stellar age (r=0.18).

The observed increase of the disk transition radius, $R_t$, and 
decrease of the mass accretion rate, $\Dot{M}_{acc}$, suggests that
the our sample disks represent an evolutionary sequence.  
Stars with different mass, spectral type and luminosity are accompanied
by disks whose characteristics seem to vary similarly with time,
i.e. they evolve in a similar way. Can we interpret this
in terms of the viscous disk model of \S~\ref{sec:modsim}?

To answer this question we can start from Eq.~\ref{eq:Rt}
and express the temporal variations of $R_t$ as:
\begin{equation}
\label{eq:R_t_res}
R_t = R_1 \left[ \frac{1}{2(2-\gamma)} \left( \frac{t}{t_s}+1 \right) \right]^{1/(2-\gamma)} 
\end{equation} 
and from Appendix~\ref{sec:macc}, the mass accretion rate 
\begin{equation}
\label{eq:macc_res}
\Dot{M}_{acc} = \frac{M_d(0)}{2(2-\gamma)t_s} \left( \frac{t}{t_s}+1 \right) 
	^{-(5/2-\gamma)/(2-\gamma)}, 
\end{equation} 
where $\gamma$ is assumed to be constant with time.
The values of $\gamma$ derived for our sample (Table~\ref{tab:res})
lead to $R_t \propto t^{0.3 - 0.8}$ and $\Dot{M}_{acc} 
\propto t^{-(1.2 - 1.4)}$. The agreement with the relation 
derived above from our observations suggests that the 
values of $R_t$ and $\Dot{M}_{acc}$ result from viscous 
evolution of disks formed with similar masses and radii 
over a time interval of about 5 Myr.

\subsubsection{Initial disk properties and time scale for the disk evolution} 

The initial disk radius, $R_1$, the initial disk mass $M_d(0)$ and 
the viscous time scale at $R_1$, $t_s$, can be estimated from Eqs. \ref{eq:R_t_res} 
and \ref{eq:macc_res}. Assuming values of $\gamma$ between -0.8 to 0.8, 
we obtain $R_1 = $25--40 AU, $M_d(0) = $ 0.05--0.4 M$_{\sun}$,
and $t_s = $0.1--0.3 Myr. In Appendix ~\ref{sec:mass} we showed that $R_1$ 
is the radius containing 63\% of the initial disk mass $M_d(0)$, 
while about 90\% of the initial disk mass $M_d(0)$ 
is contained within about $2R_1 =$ 50--80 AU.  
As noted in Appendix~\ref{sec:HR},
stellar ages increase by a factor 3-10 if B98 models are used
instead of DM97. Nevertheless, the transitional radii, $R_t$, and
the mass accretion rates, $\Dot{M}_{acc}$, still correlate with the stellar 
ages with similar correlation coefficients. $R_1$ and $M_d(0)$ are 
influenced very little by different evolutionary models but the resulting 
viscous time scales are about a factor 4 larger. In the following discussion 
we focus on deriving constraints on the processes that govern the disk 
viscosity and the disk formation from $R_1$, $M_d(0)$ and $t_s$.

\subsubsection{Implications for disk viscosity}
\label{sec:visc}
From the disk radius $R_1$, the corresponding 
viscous time scale $t_s$ and the parameter $\gamma$, we can 
derive the absolute value of the disk viscosity and its radial profile 
$\nu(R)$. From \S~\ref{sec:modsim},  $\nu(R)$ is 
expressed by 
\begin{equation}
\label{eq:visc2}
\nu(R) = \frac{1}{3(2-\gamma)^2} \frac{R_1^2}{t_s} \left( \frac{R}{R_1} 
\right)^{\gamma}.
\end{equation}
Since different values of $\gamma$ are derived from our observations 
of the millimeter dust emission, the disk viscosity is probably 
characterized by a variety of radial profiles. Thus for 
$\gamma > 0$ (DM~Tau, GO~Tau, UZ~Tau~E, GM~Aur, GSS~39 and MWC275)
the viscosity increases with radius. For $\gamma \sim 0$
(CY Tau, DN Tau, DR Tau, RY Tau, SR 24) the viscosity is virtually constant
with radius, and for the few objects with $\gamma<0$, it decreases.

\begin{figure*}
\centering
\includegraphics[angle=0,width=12cm]{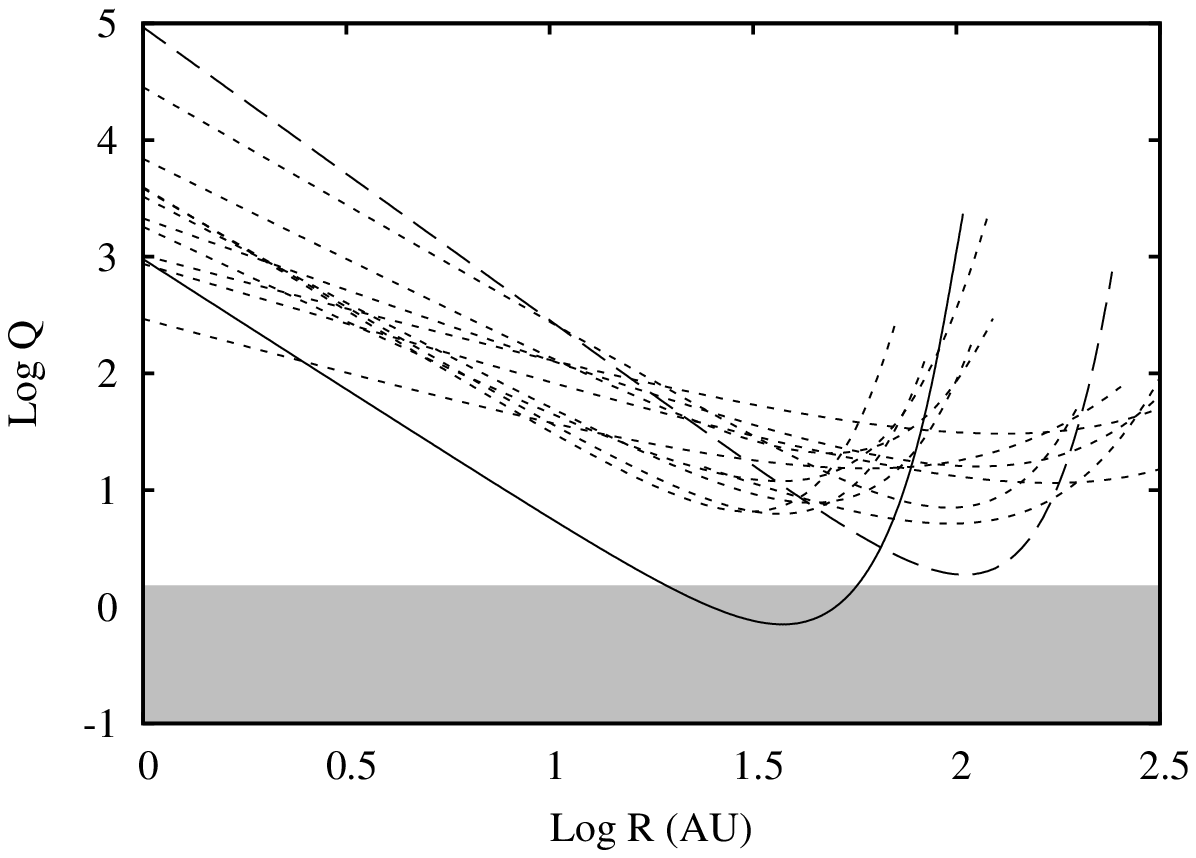}
\caption{\label{fig:Q} Radial profile of the gravitational instability parameters 
$Q$. The solid line corresponds to DG~Tau, the long dashed line to LkCa~15 and 
the short dashed lines to other sources in the sample. The grey 
region ($Q<1.5$) indicates where the disk is gravitationally unstable.}
\end{figure*}

The viscosity in a disk is generally attributed to some degree of 
turbulence that may originate from different physical processes. Two 
possible sources of turbulence are the magneto-rotational 
instability \citep{b91} and gravitational instability \citep[e.g.,][]{l04}.
Magneto-rotational instability (MRI) requires weak magnetized disks 
and a minimum ionization fraction of about $10^{-13}$ at 1~AU \citep{b00}. 
For the disk to be MRI active the maximum surface density must lie
between 10 and 100 g/cm$^2$, depending on the ionization source \citep{t08}.  
This condition is satisfied in most of our objects  (see Figure~\ref{fig:disk}).
For DG~Tau, SR~24~S and RY~Tau, however, the high density at the disk mid-plane 
inside a radius of about 30~AU probably prevents ionization. In such regions, 
where the MRI cannot operate, viscosity may originate from gravitational 
instabilities (GI). Classically GI can be parameterized by the $Q$-value (Toomre 1964) 
\begin{equation}
Q=\frac{c_s \Omega}{\pi G \Sigma(R)} 
\end{equation}
where $\Omega$ is the keplerian angular velocity 
($\Omega=\sqrt{G M_{\star}/R^3}$) and $c_s$ the sound of speed. When 
$Q \lesssim 1.5$ the disk is gravitational unstable 
and develops spiral waves to transport angular momentum outward and mass 
inward \citep{l04}. Adopting Eq.~\ref{eq:cs} for the sound speed in the disk, we can 
rewrite Q in the form 
\begin{equation}
Q \sim 230 \left( \frac{M_\star}{0.5 M_{\sun}} \right)^{1/2}
	\left( \frac{R}{10 \textrm{AU}} \right)^{-3/2} \Sigma(R)^{-1} 
T_i(R)^{1/2} 
\end{equation}
where as usual $\Sigma$ and $T_i$ are the disk surface density and 
interior temperature as in Figure~\ref{fig:disk}. 
Among the observed objects, DG~Tau has a 
gravitationally unstable disk between $\sim20$ and $\sim60$ AU 
(solid curve in Fig.~\ref{fig:Q}). For LkCa~15 (long dashed curve), Q is close 
to 1.5 between 60 and 160~AU while in all the other objects 
(short dashed curves) Q is well above the instability threshold.
Since $Q \propto \Sigma(R)^{-1}$, it strongly depends on the assumed
grain size distribution as discussed in \S~\ref{sec:dust}. In the case of
the grain size model with $q=3$ and variable $a_{max}$, $Q$ is always above the 
1.5 and gravitational instabilities cannot occur. However, the smaller surface 
density leads to an higher ionization fraction, facilitating the role of the MRI.
It seems therefore that both MRI or GI may be applicable in the observed objects.

\begin{figure*}
\centering
\includegraphics[angle=0,width=\textwidth]{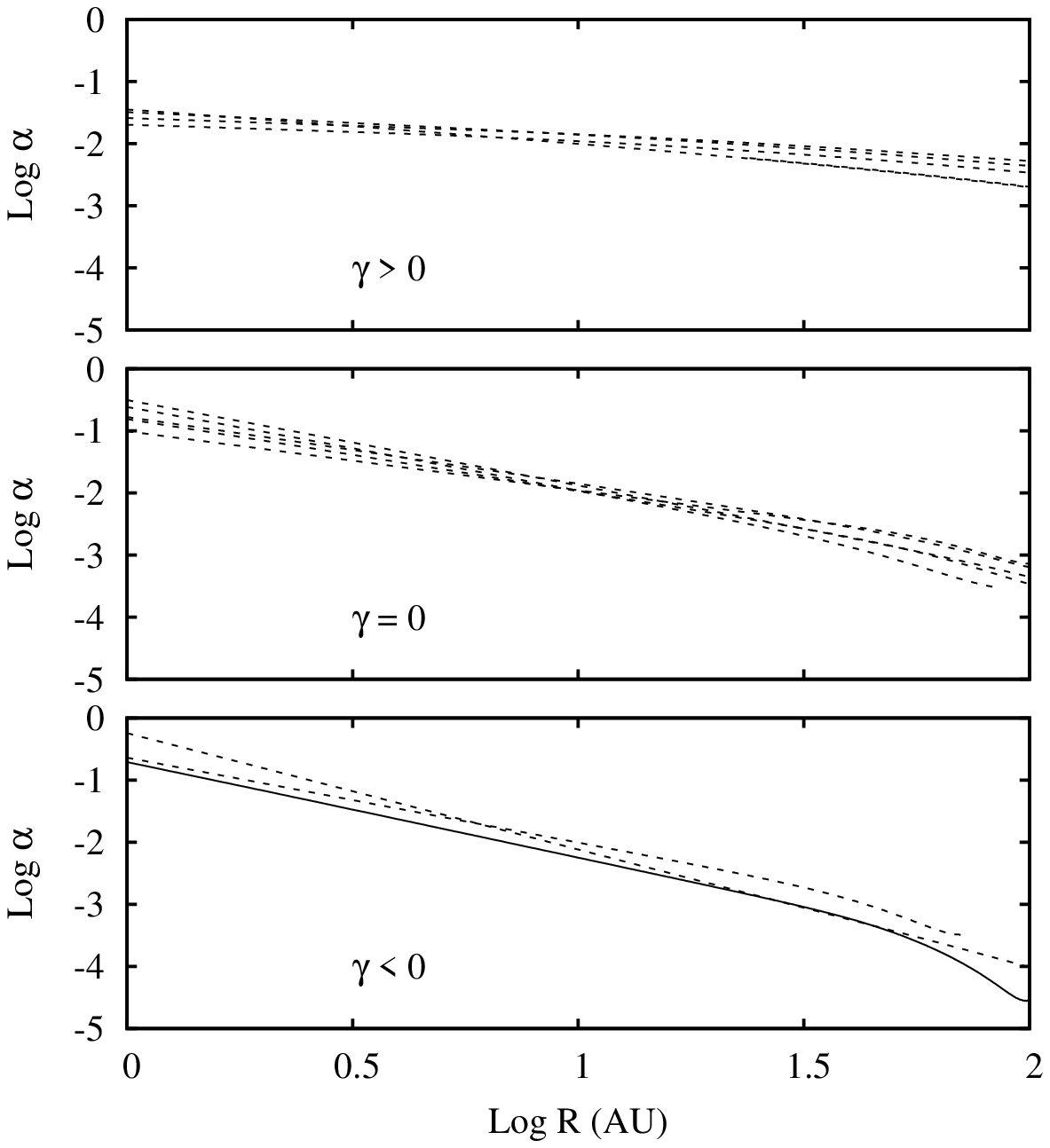}
\caption{\label{fig:alpha} Derived values of the stress parameter 
$\alpha$, for the observed circumstellar disks grouped according to  
$\gamma>0$ (upper panel), $\gamma\sim0$ (middle panel) and $\gamma<0$ (lower panel).}
\end{figure*}

We compare the viscosity profile $\nu(R)$ from Eq.~\ref{eq:visc2} with 
the theoretical expectation of MRI and GI models by adopting the classical 
$\alpha$ viscous disk parameterization (Shakura \& Sunyaev 1973) and 
writing the stress parameter $\alpha$ in the form (see Appendix~\ref{sec:alpha})
\begin{multline}
\label{eq:alpha_2}
\alpha(R) \simeq \frac{3\times10^{-2}}{(2-\gamma)^{2}} 
	\left( \frac{t_s}{1Myr} \right)^{-1}
	\left( \frac{R_1}{10\textrm{AU}} \right)^{(2-\gamma)} \\ 
	\times \left( \frac{R}{10\textrm{AU}} \right)^{(\gamma-3/2)}
	\left( \frac{M_\star}{0.5M_\sun} \right)^{1/2} T_i^{-1}(R). 
\end{multline}
It is of some importance to note that $\alpha$ is generally assumed to 
be constant in analytic modeling, although this choice has no physical 
justification and numerical simulations show variations of $\alpha$ both 
in space and time \citep{np03}. For a simple disk model that
assumes $T_i(R) \propto R^{-1/2}$
(see \S~\ref{sec:res}), constant $\alpha$ corresponds to the case $\gamma = 1$.
However, since the disk temperature $T_i$ 
deviates from the $R^{-1/2}$ profile (\S~\ref{sec:disktemp}) and 
we observe cases of $\gamma < 1$, $\alpha$ cannot be constant in the 
observed disks. Figure~\ref{fig:alpha} shows the behavior of $\alpha$ 
for $\gamma>0$ (upper panel), 
$\gamma\sim0$ (middle panel) and $\gamma < 0$ (lower panel).
In all cases, $\alpha$ decreases with the orbital radius $R$ and it 
may vary by almost 4 orders of magnitude  between 0.5 and $10^{-4}$. 
For $\gamma>0$, $\alpha$ has a shallow dependence on radius and ranges from 
about 0.03, at 1 AU, to about 0.005, at 100 AU. For $\gamma \leq 0$, 
$\alpha$ varies more rapidly and assumes values larger than 0.1 
inside radii of few AU and smaller than 0.001 outside about 30~AU.

Numerical simulations of keplerian disk affected by MRI suggest
that $\alpha$ can range from 0.005 to 0.6 
\citep[see, e.g., the review of ][]{b03}. Although detailed comparison 
of the MRI model and our observations is beyond the scope of this 
work, we note that a decrement of $\alpha$ with the radius can be achieved
for particular values of the magnetic field strength and geometry \citep{pn03}.
As discussed above, the DG Tau disk can be gravitationally
unstable between 20 and 60 AU. In this region we calculate values
of  $\alpha$  between 0.001 and 0.003 which are slightly smaller 
than $\alpha \sim 0.05$ predicted by numerical simulation
of gravitationally unstable disks \citep{l04}. 

Although our analysis is very qualitative, MRI seems to be able to 
account for the values of $\alpha$ derived in our sample disks, and, 
perhaps even the radial profiles. Nevertheless, it is still unclear whether 
circumstellar disks can be completely MRI active, or if the
MRI is effective only in the very inner part of the disk \citep[e.g.,][]{cm07}.
More importantly, it is still an open question why MRI would operate 
in different ways in our sample disks, leading to the variety of values 
of $\gamma$ and $\alpha$ discussed above.
 

\subsubsection{Implications for parent cores}
\label{sec:core}

Our viscous time scales and stellar ages suggest that the assumption
$t \gg t_s$ made in \S~\ref{sec:modsim} is probably not appropriate for 
the youngest objects, e.g., DG~Tau, DR~Tau, GSS~39 and SR~24~S. 
For these objects, with little time for disk expansion, 
the dust radial distribution should trace the initial disk structure 
resulting from the collapse of the parent core and provide 
insight into as yet unknown details of the the formation process \cite[e.g.,][]{a08}.

A simple relation between circumstellar disks and core properties can be obtained on 
the assumption that disks form from the collapse of rigidly 
rotating cores \citep{go93,s77}.
If $\omega_c$ is the core angular velocity and if the
core angular momentum $J$ is conserved during the collapse, the disk 
initial radius $R_1$ can be expressed as:
\begin{equation}
\label{eq:angular}
R_1 \simeq 25 \left( \frac{\omega_c}{10^{-14}\,\textrm{s}^{-1}} \right)^2 
	\left( \frac{M_{\star}}{1\textrm{M}_{\sun}} \right)^3 \, \textrm{AU}
\end{equation}
 \citep{hg05,d06}. Here we have assumed that (i) the disk centrifugal 
radius $R_c$ (i.e, the radius at which the angular momentum of the disk 
is equal to the angular momentum of the parent core) is similar to the 
radius that contains about 90\% of the initial disk mass (i.e, $R_c \sim 2R_1$, 
Appendix~\ref{sec:mass}), (ii) the disk mass is negligible
compared to the mass of the central star, (iii) the disk does not 
significantly expand while it is still accreting material from the envelope, 
(iv) the temperature of the core is 10 K \citep{k07,jj99} and 
(v) the magnetic field does not play a significant role in the core collapse. 

For our stellar masses (Table~\ref{tab:sample}) and disk initial radii 
between 25 and 40~AU (\S~\ref{sec:evol}), we derive core angular 
velocities between $5 \times 10^{-15}$ and $2\times 10^{-14}$ s$^{-1}$. 
Assuming simple radial profiles of the core density, we can estimate 
the specific core angular momentum $j$ (i.e., the angular momentum 
per mass unit $j=J/M$) required to form disks with initial radii in 
the observed range. We assume a core density gradient of
$\rho(r) \propto r^{-2}$ and a core radius of 0.05 pc \citep[][and references therein]{ca02}.
The core specific angular momentum, $j=(2/9) \omega_c R_{core}^2$, then 
lies between $8 \times 10^{-5}$ and $4 \times 10^{-4}$ km/s pc. 

For dense core, specific angular momenta $j=(0.5-4) \times 10^{-3}$ km/s pc 
have been derived from the measurements of velocity gradients of chemical tracer 
such as NH$_3$ and N$_2$H$^{+}$, and for cores with masses in the range 1 to 10 $M_{\sun}$ 
\citep{go93,ca02}. 
These values are an order of magnitude larger than required to form disks 
with initial radii between 25 and 40 AU, suggesting that about 10\% of
the specific core angular momentum and about 30\% of the core mass are 
conserved during the disk formation.

\subsection{Interpreting ``transitional'' disks.}
\label{sec:trans}

The surface density in disks with $\gamma<0$ (i.e., LkCa15, TW~Hya and DG~Tau) 
increases with the orbital radius and reaches a maximum at $R \sim R_t$ 
(\S~\ref{sec:res}). This effect can mimic the inner disk clearing advocated 
to explain the deficit in the near and mid-infrared excess over the 
stellar photosphere observed in ``transitional'' disks \citep{ss89} where, it is
postulated, planetary system formation may have begun \citep{es08}. 
Thus, for LkCa~15 and TW~Hya, we predict a gradual decrease of the surface density
inside radii of $\sim$60 and $\sim$17 AU respectively, in qualitative 
agreement with the radii of the dust depleted disk region already
inferred from mm-wave observations \citep{pi06,hu07}.

\begin{figure*}
     \centering	
     \label{fig:trans_a}
     \includegraphics[angle=0, width=11 cm]{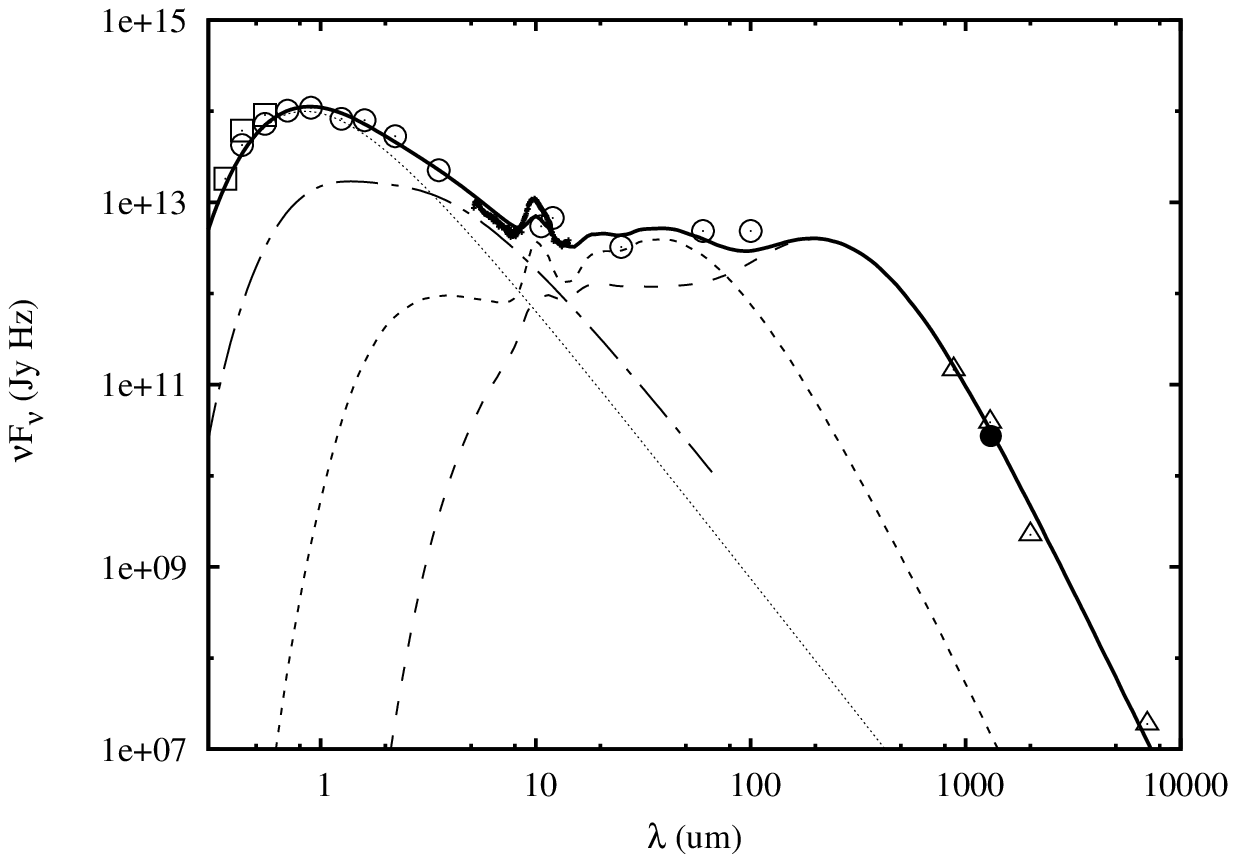}
    \caption{Spectral energy distribution of LkCa15. Flux measurements 
	from 2MASS, \citet{kh95}, \citet{kit02}, \citet{rod06}, \citet{and07} 
        are represented by open 
	squares, open circles and open triangles respectively. Our CARMA 
	data are shown as filled circle, while data between 5 and 14 $\mu$m are 
	from the Spitzer IRS archive. The solid line is the SED for the disk 
	model that fits our CARMA 1.3 mm continuum observations 
	(Tab.~\ref{tab:res}). This comprises the stellar photosphere 
	(dotted line), a ``puffed-up'' inner rim 
	(long-short dashed line), a disk surface layer (short dashed line) and 
	a disk midplane (short dashed line).}
\end{figure*}

Figure~\ref{fig:trans_a} shows the observed spectral energy distribution 
of LkCa~15 (points) and the disk model of Table~\ref{tab:res} that fits 
our 1.3 mm observations (solid line). The disk surface density is 
characterized by $\gamma = -0.8$ and
extends up to the disk inner radius without any discontinuity 
(see Figure~\ref{fig:disk_a}). At about $R=0.1$ AU, the  
dust sublimates, forming a ``puffed-up'' inner rim \citep{in05} which
emits mostly in the near-infrared between 1 and 3 $\mu$m (long-short dashed lines). 
For $R>0.1$ AU, the optically thin disk surface layer emits in the mid and far 
infrared (short-dashed line) while the colder disk midplane dominates
the emission at longer wavelengths (long-dashed line). The model reproduces
well the observed SED. 

Until now, photoevaporation, the presence of a planet, grain growth and inside-out
MRI clearing have been the main processes invoked to explain the properties 
of transitional disks \citep{al06,c05,es08,dd05,cm07}. Here we suggest 
disk viscosity as an additional means of producing a surface 
density profile and spectral energy distribution consistent with this 
class of disks. In fact, the similarity solution for the disk surface 
density predicts partially depleted inner disks whenever the viscosity 
$\nu(R) \propto R^{\gamma}$ decreases with the orbital radius. 

If transitional disks are explained by the similarity solution for the
disk surface density, the surface densities at $R<R_t$ and $R>R_t$ 
are then tightly correlated by Eq.~\ref{eq:sigma_used}. In particular, since $\gamma$ must be
negative, the surface density must fall-off very quickly for $R>R_t$ 
(see Figure~\ref{fig:sigma}).  The outer disk in transitional 
objects must be therefore characterized by a rapid decrease of the 
surface brightness measured at millimeter wavelengths. In our sample this 
hold true for LkCa15 and TW~Hya but it is not verified for DM~Tau 
and GM~Aur. For the latter two objects, the observed dust continuum emission 
at $R \gtrsim 45$ AU leads to positive value of $\gamma$, implying 
that the surface density in DM~Tau and GM~Aur increases smoothly up to the inner 
disk radius (see Figure~\ref{fig:disk}). This contrasts with the observations 
of dust depleted inner disks within about 7 and 20 AU \citep{c05}.

We suggest that this interpretation of ``transitional disks'' is applicable
when the dust-depleted inner region occupies a significant fraction 
of the disk extent. In LkCa15 and TW Hya, $R_t$ is in fact more 
than 20\% of the disk radius $R_d$ (Table \ref{tab:res}). By contrast, 
when the dust-depleted inner region is only a few percent of the disk 
radius, as for DM~Tau and GM~Aur, the presence of a planet, the 
inside-out MRI clearing or the photoevaporation by the central star 
are more probable explanations for the inner disk clearing.

\subsection{Similarity with the solar nebula}
\label{sec:solar}

A fundamental question regarding pre-main sequence disks is whether
they will evolve into planetary systems similar to our own. Addressing
this question is difficult mainly because we have a limited knowledge
on the properties of the solar nebula. 
A recent re-analysis of the distribution of solid bodies in the Solar
system by \citet{d05} differs from earlier studies \citep{k70,w77,hay81}
and suggests that the surface density distribution
in the solar nebula at $10^5$--$10^6$ yr from its formation is described by
\begin{equation}
\label{eq:davis}
\Sigma(R) \simeq 1.14\times10^{3} R^{-1/2} e^{-0.024 R^{3/2}}.
\end{equation} 
The numerical constants are chosen to recover the total 
mass of the solar nebula, 0.02 $M_\sun$ \citep{k70}, 
and its specific angular momentum $j=8.7\times10^{-6}$ km/sec pc \citep{c00}. 

The solar nebula surface density of Eq.~\ref{eq:davis} is based on the
assumption that $\gamma=0.5$, and corresponds to the similarity solution 
expressed by Eq.~\ref{eq:sigma_used} for $R_t = 6$~AU and 
$\Sigma_t = 340$ g/cm$^2$. The total mass for the solar nebula 
are quite similar to the values measured for GSS~39, GM~Aur, DM~Tau and GO~Tau. 
However, the solar nebula transition radius, $R_t$, is at least a factor 
of 4 smaller than what found for objects with ages of $10^5$--$10^6$ yr
(see \S~\ref{sec:prof} and \S~\ref{sec:core}). 
 
Indeed, among our disk sample, only TW~Hya may be a good match to the solar nebula,
being the only old object characterized by a small transition radius. 
If TW~Hya underwent viscous evolution similar to that discussed in \S~\ref{sec:evol},  
it probably formed with an initial transition radius of 4-8~AU, similar to that
predicted for the solar nebula.

\section{Conclusions}
\label{sec:conc}
We presented high angular resolution (0.7$\arcsec$) interferometric
observations of the 1.3 mm continuum emission from 14 pre-main sequence 
circumstellar disks. The disk surface brightness is characterized by
a range of radial profiles. Adopting the similarity solution 
for disk surface density \citep{har98} and a self consistent 
disk emission model \citep{is07}, we derived for each disk a
surface density radial profile defined by the transition radius $R_t$,
the surface density $\Sigma_t$ and the slope of the disk viscosity $\gamma$, 
as well as a radial temperature profile and its orientation in space.
Assuming a constant dust opacity throughout the disk, we find that 
the different surface brightness profiles are mainly due to differences 
in $R_t$ and $\gamma$, which in turn imply different disk surface densities.

From a comparison of the disk surface density and the stellar properties, it 
appears that the disk transition radius $R_t$ is correlated with 
the stellar age and increases from $\sim$20 to $\sim100$ AU over about 3-5 Myr. 
This disk expansion appears to be accompanied by a decrease in the mass 
accretion rate. We argue that these temporal variations of the disk radius 
and the mass accretion rate support a scenario in which disks form 
an evolutionary sequence. The observed evolution is in 
qualitative agreement with that of viscous disk models if 
the initial disk masses are between 0.05 and 0.4 M$_{\sun}$ 
and the initial disk radii range from 20 to 45 AU. Note however that
disk masses can vary by one order of magnitude depending on the
assumed dust opacity. The temporal 
variation of the disk radius and mass constrains the viscous time 
scale $t_s$ at the disk initial radius $R_1$ to be between 0.1 to 0.3 Myr.

The viscous disk model assumes that the disk viscosity $\nu(R)$ scales 
as $R^{\gamma}$. Among our sample, $\gamma$ ranges 
from -0.8 to 0.8, leading to a large variety of viscosity radial profiles.  
Parameterizing the disk viscosity in terms of the stress parameter $\alpha$, 
we show that $\alpha$ scales with radius $R$ roughly as $R^{\gamma-1}$. 
Since $\gamma$ is always smaller than 1, $\alpha$ must
decrease with the orbital radius. We suggest that $\alpha$ may vary by almost 4 
orders of magnitude between 0.5 and $10^{-4}$. These values are in general  
agreement with MRI models, which is probably the main source of viscosity at the
surface densities of our disks. However, its still an open question why MRI should
operate differently among our sample, leading to a large range of values of $\gamma$.  

The ages of the younger stars in our sample are comparable to the 
viscous time scale of 0.1-0.3 Myr. It seems likely that, with little time
for the disk expansion, the dust radial distribution should trace the initial 
disk structure resulting from the collapse of the parent core.
For typical assumptions on the core rotation, radial density profile and radius, 
we derive the core specific angular momentum $j=J/M$ required to form disks with 
initial radii between 25 to 40 AU. We argue that $j$ must range from $8 \times 10^{-5}$ 
to $4 \times 10^{-4}$ km/s pc, and suggest that this corresponds to about 10\% of 
the specific angular momentum measured in dense cores. It seems therefore 
likely that during the star formation process, about 10\% of the core angular 
momentum is transferred to the circumstellar disk. Alternatively, 
10\% efficiency in conserving angular momentum implies that about 30\% 
of the core mass is used to form the central star.
We believe that the attempt to correlate the properties of circumstellar 
disks and dense cores, though still very qualitative, is an important step forward 
to understand the role of disks in conserving the angular momentum during 
the star formation process. Clearly, the capability to investigate this aspect 
in detail are actually hampered by the sensitivity and resolution of the 
existing interferometers, which limit the analysis to a few bright objects. 
However, the ongoing improvement of facilities such as CARMA and VLA, 
and the advent of ALMA, will enable us to expand the number of spatially 
resolved disks by a large factor and have a more complete view on the relations 
between dense core and young disks.

We point out that in disks with $\gamma <0$, the surface density $\Sigma(R)$
increases with radius $R$ and reach a maximum at about the transitional radius 
$R_t$. We argue that this particular behavior of $\Sigma(R)$ can mimic the inner disk 
clearing advocated to explain the dominant characteristic of some ``transitional disks'', 
namely a deficit in the near and mid-infrared excess over the stellar photosphere and 
the presence of an ``hole'' in the surface brightness observed at millimeter wavelengths.
For LkCa15, we show that a surface density with $\gamma = -0.8$ that extends 
without any discontinuity up to the disk inner radius located at few stellar radii can 
reproduce both the SED and the 1.3 mm continuum emission. By contrast, we find no clear
explanation in terms of the similarity solution of the surface density for 
the dust depleted inner disks around DM~Tau and GM~Aur. It seems likely that 
``transitional disks'' may originate from a large variety of effects. 

Finally, it appears that most of the disks in our sample are very different 
from the currently-accepted view of the solar nebula. While most of them have 
masses similar to the minimum mass solar nebula, 0.02 M$_{\sun}$, their 
transitional radii are at least a factor 4 larger 
than the value $R_t \sim 6$ AU derived from the actual distribution of solid bodies 
in the Solar system \citep{d05}. The exception is TW~Hya which has a very small 
disk radius, $R_t \sim 17$ AU, compared to its large age, $t=7$ Myr. We argue that 
TW~Hya disks probably formed with $R_t$ in the range from 4 to 8 AU, and may well 
reflect the properties of our early solar system.

\acknowledgments
We are indebted to Meredith Hughes and David Wilner for providing
the SMA data of GM~Aur and TW~Hya. We thank Antonella Natta, Sean Andrews
and Leonardo Testi for the useful discussions and the referee for 
the very useful comments. We thank the OVRO/CARMA staff 
and the CARMA observers for their assistance in obtaining the data. 
We acknowledge support from the Owens Valley Radio Observatory, 
which is supported by the National Science Foundation through grant AST 05-40399.
This work was performed in part under contract with the Jet Propulsion Laboratory 
(JPL) funded by NASA through the Michelson Fellowship Program. JPL is 
managed for NASA by the California Institute of Technology.






\appendix 

\section{Stellar ages and masses}
\label{sec:HR}
\begin{figure*}[t]
\begin{center}
	\includegraphics[angle=0,scale=0.7, bb=30 173 550 550]{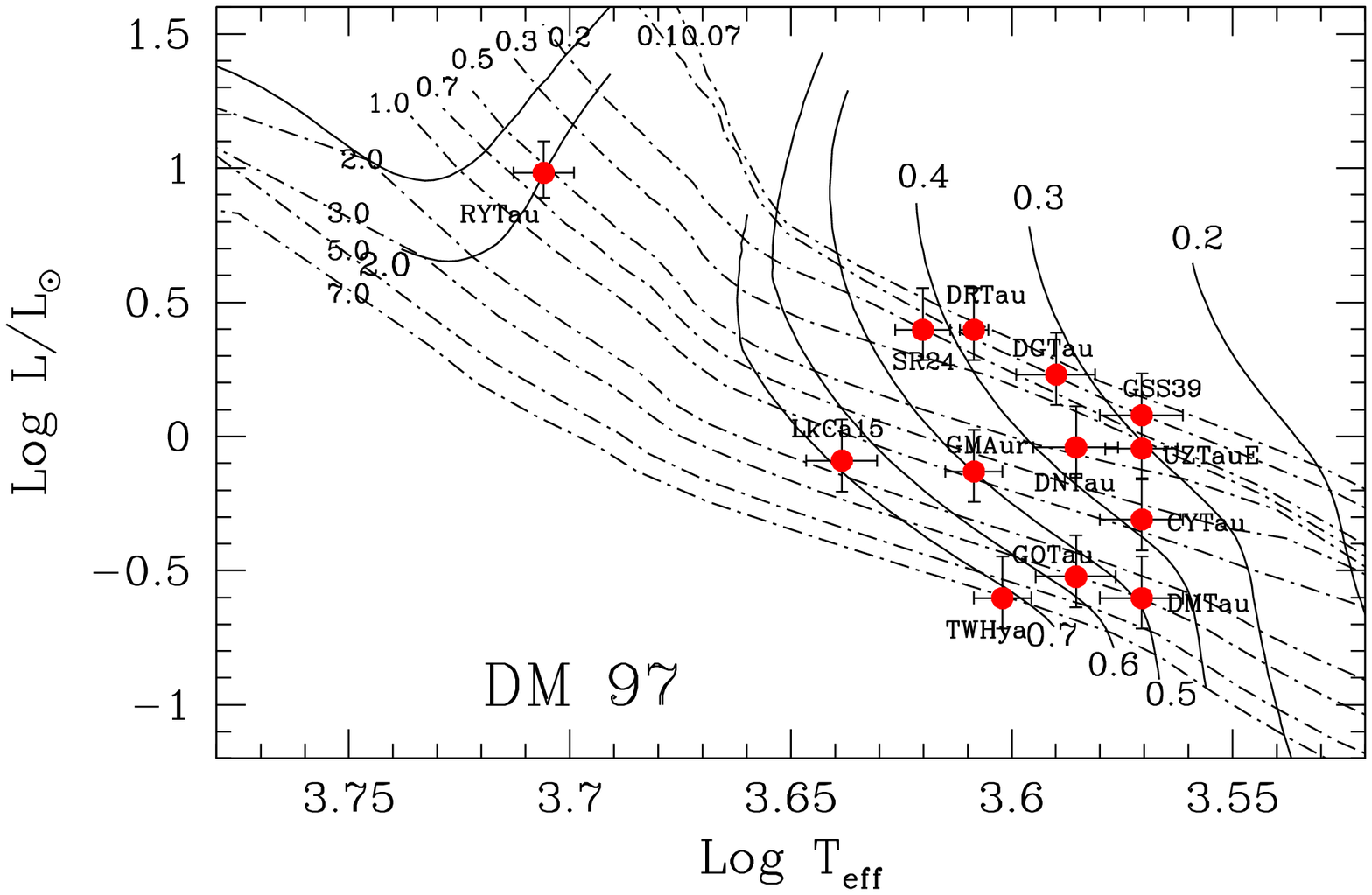}
	\includegraphics[angle=0,scale=0.7, bb=30 173 550 550]{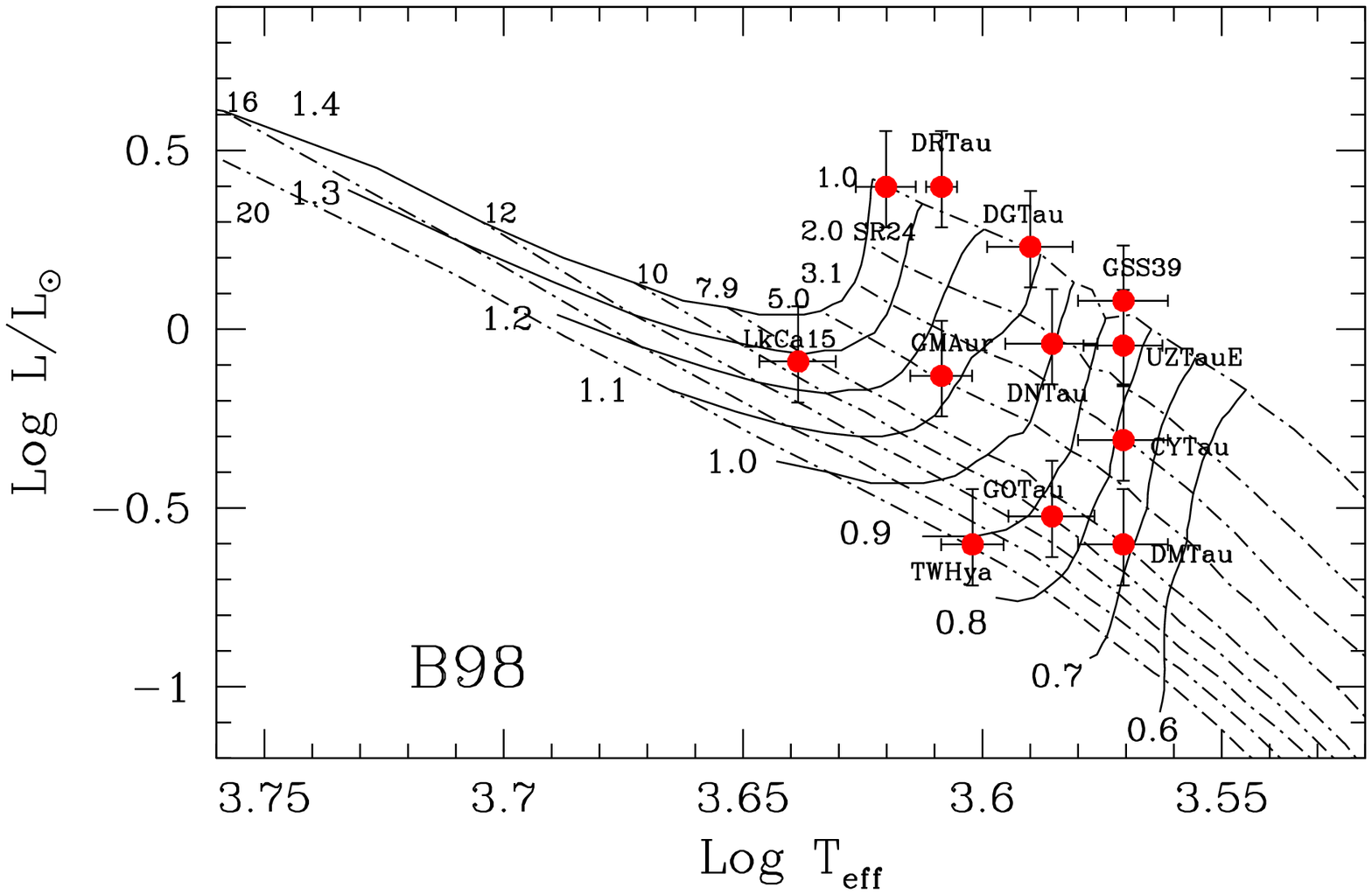}
	\caption{Position on the H-R diagram of the observed sources. The upper 
	panel show the theoretical models by D'Antona \& Mazzitelli (1997) 
	while the lower panel the models of Baraffe et al. (1998). The dashed lines 
	correspond to the stellar isocrones for ages in Myr as labeled at the 
	left end of the lines while the solid lines correspond to the stellar evolution 
	sequence for stellar masses between 0.2 and 1.7 $M_{\sun}$ as labeled at the 
	lower end of the lines.}
\label{fig:HR}
\end{center}
\end{figure*}

\begin{figure*}[t]
\begin{center}
	\includegraphics[angle=0,scale=0.8]{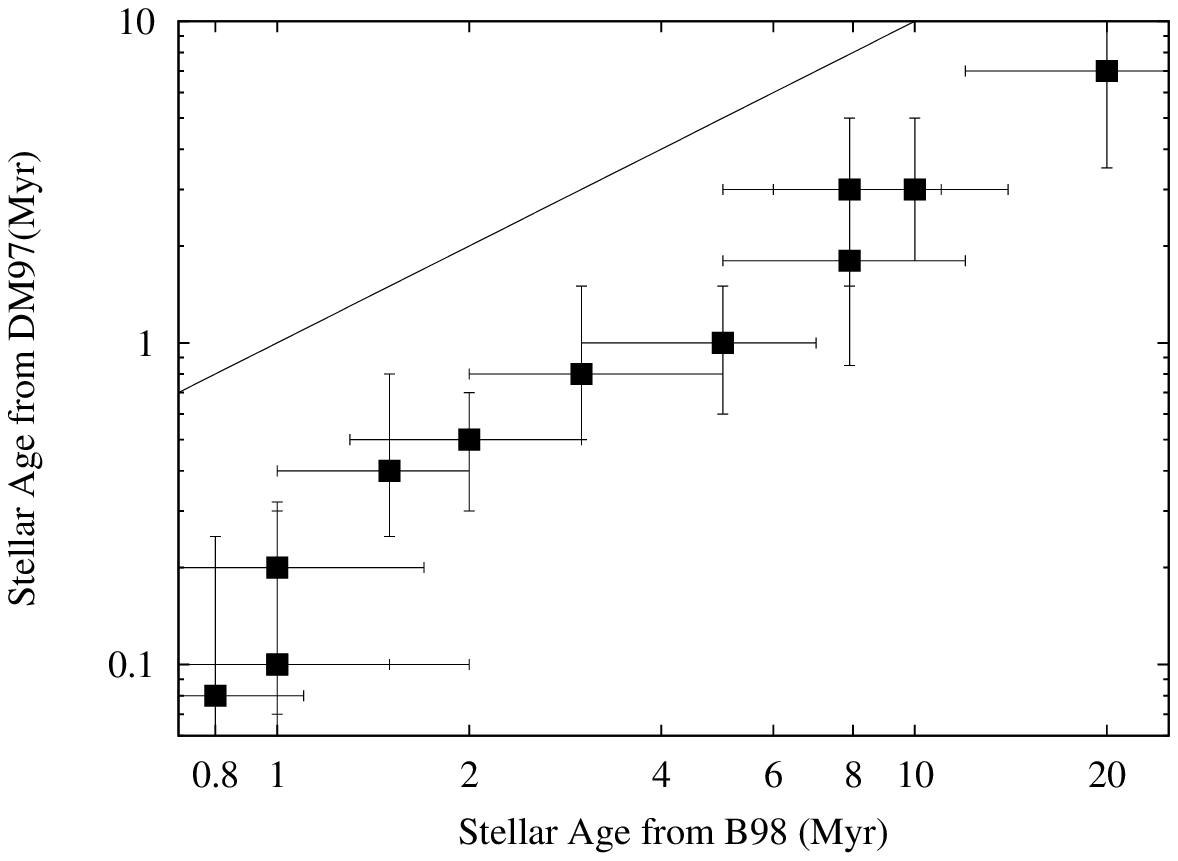}
	\includegraphics[angle=0,scale=0.8]{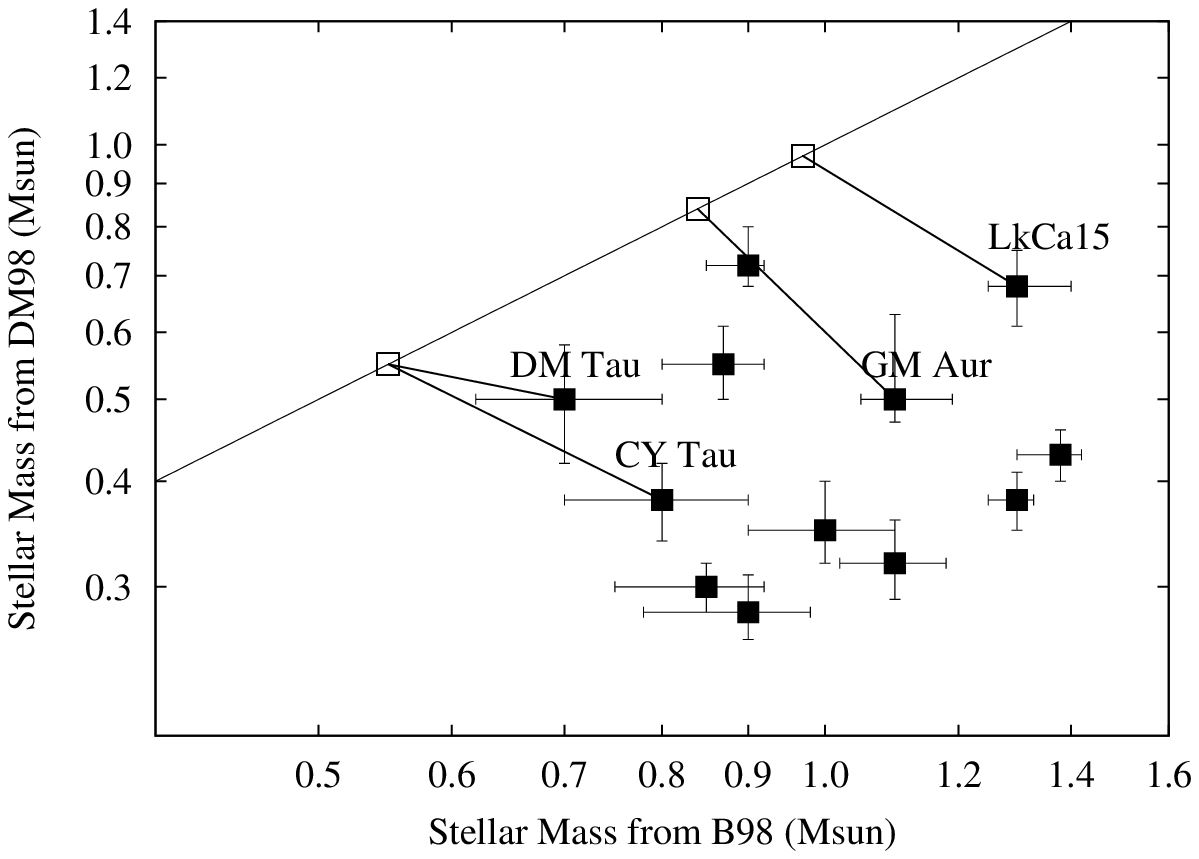}
	\caption{Upper panel: stellar ages derived from DM97 versus the 
	ages derived from the B98 models. The solid lines correspond to equal 
	ages. The comparison between stellar masses is shows in the lower panel.
	The open squares identify the dynamical stellar masses derived by Simon 
	et al. (2000) for DM~Tau, CY~Tau, GM~Aur and LkCa~15}
	\label{fig:comp_HR}
\end{center}
\end{figure*}

Figure~\ref{fig:HR} shows the location of our sample stars in the H-R 
diagram with the evolutionary models of \citet[][hereafter DM97]{dm97} 
and \citet[][B98]{bar98}. The uncertainties on the stellar temperature 
correspond to half a spectral type, 
while the stellar luminosity is uncertain by 30\%.
From DM97 models, the stellar ages 
range from ~0.1~Myr (DR~Tau and DG~Tau) to 7~Myr (TW~Hya). 
In Taurus-Auriga alone the age spread is 1--3 Myr. 
However, as illustrated in the upper panel of Figure \ref{fig:comp_HR}, these ages are 
smaller than those inferred from B98 by factor of 2.6-10.  Similarly, the lower panel of 
Figure~\ref{fig:comp_HR} shows that the stellar masses inferred
from B98 are systematically larger than those from  DM97 by factors of 1.2-3.4.
Neither model reproduces the dynamical masses derived by \citet{sim00} 
for DM~Tau, GM~Aur, CY~Tau and LkCa~15.
Typically the DM97-derived masses are lower than the dynamical masses 
by 10-40\% while B98-masses are lower by 20-30\%. 
For the purpose of our analysis we have adopted DM97 models because they
enable us to derive masses and ages for all the stars in our sample, while
RY~Tau and MWC~275 are outside the temperature and luminosity range of the BH98 model.

\section{Fitting process}
\label{sec:MCMC} 

For fixed stellar parameters, dust opacity and disk inner 
radius, the dust emission model is defined by the state ${\mitbf x} 
\{i,PA,R_t,\Sigma_t,\gamma\}$ where $i$ is the disk inclination, PA is 
the disk position angle measured from North to East, $R_t$ is the disk 
transition radius, $\Sigma_t$ is the disk surface density at $R_t$ and 
$\gamma$ defines the shape of the disk surface density as discussed in 
\S~\ref{sec:modsim}. The model that best fits the observations is found 
through the minimization of $\chi^2$ (Eq.~\ref{eq:chi2}).
We adopt a Bayesian approach in which the joint 
probabilities for the observed data {\boldit d} and the given model 
state {\boldit x} are described as the product of the probability 
of the observed data {\boldit d} given the model parameters {\boldit x} 
(i.e., the likelihood), and a known prior probability distribution function 
$p({\mitbf x})$ of the model parameters:
\begin{equation}
\label{eq:bayes}
p({\mitbf x}|{\mitbf d}) \propto p({\mitbf x}) p({\mitbf d}| {\mitbf x}).
\end{equation}
In this framework, the best-fit model corresponds to the state
{\boldit x} that maximizes the a posteriori distribution 
$p({\mitbf x}|{\mitbf d})$.

To characterize the posteriori distribution we adopt a Markov Chain Monte 
Carlo (MCMC) method (see, e.g., Ford 2005, Fitzgerald et al.~2007). 
This method generates a chain of states {\boldit x} sampled from a
desired probability function $p({\mitbf x})$, whose equilibrium distribution is 
equal to the posteriori distribution $p({\mitbf x}|{\mitbf d})$.
In our specific case, since complex visibilities errors  
are described by a normal distribution (Wrobel \& Walker 1999), the 
probability of the observables (i.e, the real and imaginary part of the complex
visibility) given the model state {\boldit x}, is expressed by the 
$\chi^2$ distribution. To a first approximation this is proportional 
to  $e^{-\chi^2({\mitbf x})/2}$. In the most general case, where 
all model states have the same prior probability and 
$p({\mitbf x}) \sim 1$, the posteriori joint probability 
is also roughly proportional to $e^{-\chi^2({\mitbf x})/2}$. The state that 
maximizes $p({\mitbf x}|{\mitbf d})$ therefore correspond 
to the state that minimizes $\chi^2({\mitbf x})$, justifying 
the choice of the $\chi^2$ as the maximum likelihood estimator. 

The MCMC fitting is realized following \citet{f05} and is briefly 
summarized here. First, the chain is constructed using the Metropolis-Hastings 
algorithm with the Gibbs sampler, and generating a chain of models where 
the (n+1)-th state depends only on the n-th state through a specified transition 
probability $q(\textrm{n}|\textrm{n+1})$. In practice, at each n-th step of 
the chain we (1) generate a random trail state {\boldit x\arcmin} 
adopting a transition probability described by a Gaussian distribution 
centered on {\boldit x}, (2) calculate $\chi^2({\mitbf x'})$, and (3) 
accept the trail state as the new (n+1)-th state of the chain if it 
satisfies an acceptance probability $\alpha({\mitbf x'}|{\mitbf x_n})$ 
specified by the M-H algorithm, namely if
\begin{equation}
\alpha({\mitbf x'}|{\mitbf x_n})=min \left\{ -\frac{1}{2}[\chi^2({\mitbf x'}) - \chi^2({\mitbf x_n})],1 \right\} \geq u
\end{equation} 
where $u$ is a random number generated from a uniform distribution between 0 and 1.
For the Gibbs sampler, the trial state {\boldit x\arcmin} is generated by substituting 
only a subset of parameters from the state {\boldit x}. In particular, since $R_t$, 
$\Sigma_t$ and $\gamma$ are correlated (Eq.~\ref{eq:sigma_used}), we choose to 
change these three parameters at the same time, while varying
the inclination and the position angle independently. At each step, the 
parameter(s) to be updated ($x_\mu$) are then randomly modified with a 
transition probability of
\begin{equation}
q(x'_\mu|x_\mu) = \frac{1}{\sqrt{2\pi\beta_\mu^2}} \exp{\left[ 
-\frac{(x'_\mu - x_\mu)^2}{2\beta^2_\mu} \right]},
\end{equation}
where the variance $\beta_\mu$ defines the variability interval for each parameter.
The model fitting results depend neither on the choice of the transition 
probability $q({\mitbf x'}|{\mitbf x})$ nor on how the Gibbs sampler is implemented. 
However these strongly effect the efficiency of the model fitting and 
must be carefully chosen to allow a fast convergence toward the equilibrium 
distribution. Here, each parameter is allowed to vary over a large 
interval and the variance $\beta_\mu$ is chosen so that the overall acceptance
rate is close to the optimal value of $\sim$0.25 (Gelman et al. 2003). 

To select the initial state of the chain we adopt two different approaches. 
First, the initial parameters for the disk are set based on the Gaussian 
fit to the observations (Table~\ref{tab:gau_fit}); the initial inclination 
and PA are calculated from the aspect ratio of the emission, the
transition radius $R_t$ is assumed to be the half width along 
the major axis at half maximum, $p$ is 
set to be 0.5 and $\Sigma_t$ is randomly chosen between 0.1 and 1000 g/cm$^2$. 
This choice is usually close to what provided by the best fit model and 
enables a quick convergence of the chain. Unfortunately, like most algorithms 
developed to minimize the $\chi^2$, the MCMC can be trapped in local minima 
if they are separated by sufficiently high $\chi^2$ barriers. To compensate,
additional randomly initialized MCMC chains were run to verify that 
they all converge to the same final state.       

Once the equilibrium distribution of the MCMC chain is sufficiently well 
sampled -- usually requiring a run of a minimum of 10$^4$ models -- the 
distribution of each parameter is obtained through marginalization, i.e. by 
integrating the posteriori distribution (now equivalent to the MCMC 
equilibrium distribution) over all the parameters except the one in which 
we are interested. We show in Figure~\ref{fig:prob}
the obtained posteriori distributions for the model parameters. These 
are generally consistent with a normal profile, supporting our decision
to adopt a Gaussian transition probability $q({\mitbf x'}|{\mitbf x})$, 
and enables expressing the parameter uncertainties in terms of the 
standard deviation $\sigma$ of the probability distribution. 

\section{Cumulative and total disk mass}
\label{sec:mass}
We integrate the surface density (Eq.~\ref{eq:sigma_used}) to derive 
the cumulative disk mass, i.e., the disk mass contained within a radius 
$R$ at time $t$. Since $R_{in} \ll R_t$ we obtain
\begin{equation}
\label{eq:cummass}
M_d(R,t) = 4\pi\Sigma_t R_t^2 e^{1/2(2-\gamma)} \\
\times  \left\{ 1 - \exp{\left[ -\frac{1}{2(2-\gamma)} 
\left( \frac{R}{R_t} \right)^{(2-\gamma)}  \right] }   \right\}
\end{equation}
The total disk mass at the time $t$ is therefore obtained for 
$R \rightarrow \infty$ in the form
\begin{equation}
\label{eq:mdisk}
M_d(t) =  4\pi\Sigma_t R_t^2 e^{1/2(2-\gamma)}.
\end{equation}
Using the last two equations we derive that the disk mass contained 
inside the transition radius $R_t$ 
is 39\% of the total disk mass if $\gamma=1$ and 22\% if $\gamma=0$.

Using Eq.~\ref{eq:cummass} and Eq.~\ref{eq:mdisk} we can also 
demonstrate that $R_1$ is the radius containing
$\sim$63\% of the initial disk mass $M_d(t=0)$. We can write
\begin{equation}
\label{eq:m0}
\frac{M_d(R_1,0)}{M_d(0)}= 1 - \exp\left[ -\frac{1}{2(2-\gamma)} \left( \frac{R_1}{R_{t,0}}\right)^{(2-\gamma)} 
\right]
\end{equation}  
where the transitional radius at $t=0$, given by Eq.~\ref{eq:Rt}, is
\begin{equation}
\label{eq:Rt0}
R_{t,0}=R_1\left[ \frac{1}{2(2-\gamma)} \right]^{1/(2-\gamma)}.
\end{equation}
Substituting this latter equations in Eq.~\ref{eq:m0} we obtain
\begin{equation}
\frac{M_d(R_1,0)}{M_d(0)}= 1 - e^{-1} \simeq 0.63
\end{equation}

We finally calculate that about 90\% of the initial disk mass is contained
within $2R_1$. From Eq.~\ref{eq:m0} and Eq.~\ref{eq:Rt0} we can write
\begin{equation}
\frac{M_d(2R_1,0)}{M_d(0)}= 1 - e^{-2^{(2-\gamma)}} 
\end{equation}
which is equal to 0.86 for $\gamma=1$ and 0.98 for $\gamma=0$.


\section{Mass accretion rate on the central star}
\label{sec:macc}
From Eq.~\ref{eq:sigma} and \ref{eq:sigma_used} the surface density 
$\Sigma_t$ at the transition radius $R_t$ can be expressed as:
\begin{equation}
\label{eq:E1}
\Sigma_t = \frac{C}{3\pi \nu_1}T^{-(5/2-\gamma)/(2-\gamma)} \left( \frac{R_1}{R_t}\right)^\gamma \times
\exp{\left[-\frac{1}{2(2-\gamma)} \right]}
\end{equation}

Writing the initial disk mass $M_d(0)$ in the form
\begin{equation}
\label{eq:m1}
M_d(0)=\frac{2}{3} \frac{C}{\nu_1} \frac{R_1^2}{2-\gamma}
\end{equation}
we can eliminate the ratio $C/\nu_1$ from Eq.~\ref{eq:E1} to obtain 
\begin{equation}
\label{eq:sigma_t_2}
\Sigma_t = \frac{M_d(0)}{4\pi R_t^2} T^{-1/2(2-\gamma)} e^{-1/2(2-\gamma)}.
\end{equation} 
Substituting the definition of $R_t$ from Eq.~\ref{eq:Rt},  $\Sigma_t$  takes the form 
\begin{equation}
\Sigma_t \propto \frac{M_d(0)}{4\pi R_1^2} T^{-5/(2(2-\gamma))}.
\end{equation}

Starting from the expression of the mass flow \citep{har98} 
\begin{equation}
\Dot{M}(r,t)=C T^{-\frac{(5/2-\gamma)}{(2-\gamma)}} \exp{\left[-\frac{r^
{(2-\gamma)}}{T} \right]} \times \left[ 1- \frac{2(2-\gamma)r^{(2-\gamma)
}}{T} \right]
\end{equation}
and substituting the expression for the constant $C$ 
derived from Eq.~\ref{eq:m1} we can express the mass flow as
\begin{equation}
\label{eq:Md}
\Dot{M}(R,t)=\frac{M_d(0)}{2(2-\gamma)t_s} \left[ 1 - \left( \frac{R}{R_t} \right)^{(2-\gamma)} \right] 
	   T^{-(5/2-\gamma)/(2-\gamma)} \exp{ \left[ -\frac{R^{(2-\gamma)}}{2(2-\gamma)R_t^{2-\gamma}} \right] }
	 ,
\end{equation}
where $t_s$ is the viscous time scale defined by the Eq.~\ref{eq:ts} and 
\begin{equation}
r = \frac{R}{R_1} = \frac{R}{R_t} \frac{R_t}{R_1} = \frac{R}{R_t}  \left[ \frac{T}{2(2-\gamma)} \right]^{1/(2-\gamma)}.
\end{equation}
Note that for $\gamma=1$, Eq.~\ref{eq:Md} reduces to Eq.~35 in \citet{har98}.

At radii smaller than the transitional radius $R_t$, the material within 
the disk moves inward and finally falls onto the central star. The mass 
accretion rate $\Dot{M}_{acc}(t)$ is given by Eq.~\ref{eq:Md} 
with $R$ equal to the radius at which the disk is truncated by the accretion 
process. Since this radius is of the order of a fraction of AU and
much smaller than the transitional radius $R_t$, $\Dot{M}_{acc}$ takes the form
\begin{equation}
\label{eq:macc}
\Dot{M}_{acc}(t)=\frac{M_d(0)}{2(2-\gamma)t_s} T^{-(5/2-\gamma)/(2-\gamma)}.
\end{equation} 


\section{Derivation of $\alpha$ in the case $\gamma \neq 1$}
\label{sec:alpha}
In this section we derive the value of $\alpha$ for the similarity
solution of the disk surface density in the general case $\gamma \neq 1$.
We adopt the classical $\alpha$ parameterization of the disk viscosity
in the form (Shakura \& Sunyaev 1973)
\begin{equation}	
\label{eq:alpha}	
\nu = \alpha c_s H.	
\end{equation} 
If the disk is in keplerian rotation, vertically isothermal 
and in hydrostatic equilibrium (see the discussion in \S~\ref{sec:mod}), 
the sound speed is given by
\begin{equation}
\label{eq:cs}
c_s = H \cdot \Omega = H \cdot \sqrt{GM_{\star}/R^3}
\end{equation}  
with
\begin{equation}
\label{eq:H}
H = R^{3/2} \sqrt{\frac{k_b T_i(R)}{\mu m_H G M_{\star}}},
\end{equation}
where $k_b$ is the Boltzman's constant, $\mu = 2.33$ g mol$^{-1}$ 
(Ruden and Pollack 1991) is the mean molecular weight of the circumstellar 
material, $m_H$ is the proton mass, $G$ is the gravitational constant 
and $T_i(R)$ is the disk interior temperature. Substituting Eq.~\ref{eq:cs} 
and \ref{eq:H} in Eq.~\ref{eq:alpha}, we can 
express the disk viscosity as a function of the radius in the form
\begin{equation}
\label{eq:visc3}
\nu(R) =  \frac{k_b}{\mu m_H \sqrt{G M_{\star}}} \cdot \alpha R^{3/2} T_i(R). 
\end{equation}
The comparison of Eq.~\ref{eq:visc2} and Eq.~\ref{eq:visc3} provides therefore
an expression of $\alpha$ in the form
\begin{equation}
\alpha(R) \simeq \frac{3\times10^{-2}}{(2-\gamma)^{2}} 
	\left( \frac{t_s}{1Myr} \right)^{-1}
	\left( \frac{R_1}{10\textrm{AU}} \right)^{(2-\gamma)} 
	\left( \frac{R}{10\textrm{AU}} \right)^{(\gamma-3/2)}
	\left( \frac{M_\star}{0.5M_\sun} \right)^{1/2} T_i^{-1}(R). 
\end{equation}.


\begin{deluxetable}{llllllllllll}
\rotate
\tablecaption{Sample properties \label{tab:sample}}
\tablewidth{0pt}
\tablehead{
\colhead{Object} & \colhead{$\alpha$(2000)} & \colhead{$\delta$(2000)} & \colhead{ST} & \colhead{L$_\star$} & \colhead{T$_\star$}  & \colhead{$log(L_{acc})$} & \colhead{Ref.} & \colhead{R$_{\star}$} &
\colhead{$log(\Dot{M}_{acc})$} & \colhead{M$_\star$} & \colhead{Age} \\
\colhead{(1)} & \colhead{(2)} & \colhead{(3)} & \colhead{(4)} & \colhead{(5)} & \colhead{(6)} & \colhead{(7)} & \colhead{(8)} & \colhead{(9)} & \colhead{(10)} & \colhead{(11)} & \colhead{(12)}
}
\startdata														    
CY~Tau   & 04:17:33.73  & 28:20:46.95 & M1 & 0.47 & 3720 & -1.34 & {\footnotesize 1}         & 1.68 & -8.52 & 0.4 & 0.8 \\ 
DG~Tau   & 04:27:04.70  & 26:06:16.39 & M0 & 1.70 & 3890 &  0.70 & {\footnotesize 7}         & 2.87 & -6.39 & 0.3 & 0.1 \\ 
DM~Tau   & 04:33:48.73  & 18:10:09.96 & M1 & 0.25 & 3720 & -1.05 & {\footnotesize 1}         & 1.20 & -8.32 & 0.5 & 3.0 \\ 
DN~Tau 	 & 04:35:27.37  & 24:14:58.90 & M0 & 0.91 & 3850 & -1.80 & {\footnotesize 1}         & 2.14 & -8.97 & 0.4 & 0.5 \\ 
DR~Tau   & 04:47:06.22  & 16:58:42.87 & K7 & 3.00 & 4060 &  0.44 & {\footnotesize 3,6,7}     & 3.19 & -6.68 & 0.4 & 0.1 \\ 
GO~Tau   & 04:43:03.09  & 25:20:18.59 & M0 & 0.28 & 3850 & -0.98 & {\footnotesize 1}         & 1.23 & -8.33 & 0.6 & 3.0 \\ 
LkCa15   & 04:39:17.78  & 22:21:03.52 & K5 & 0.74 & 4350 & -1.75 & {\footnotesize 1}         & 1.60 & -9.17 & 0.7 & 1.8 \\ 
RY~Tau   & 04:21:57.41  & 28:26:35.56 & K1 & 7.60 & 5080 &  0.20 & {\footnotesize 5}         &  2.92 & -7.11 & 2.0 & 0.5 \\ 
UZ~Tau~E & 04:32:43.07  & 25:52:31.14 & M1 & 0.90 & 3720 &  0.17 & {\footnotesize 5}         & 2.28 & -6.90 & 0.3 & 0.4 \\ 
GM~Aur	  & 04:55:10.98	& 30:21:59.38  & K7 & 0.74 & 4060 & -1.15 & {\footnotesize 1}        & 1.4  & -8.55 & 0.5 & 1.0 \\ 
GSS~39   & 16:26:45.00  &-24:23:07.70 & M1 & 1.20 & 3720 & -0.40 & {\footnotesize 2}         & 2.64 & -7.43 & 0.3 & 0.1 \\ 
SR~24~S    & 16:26:58.50  &-24:45:36.90 & K6 & 2.50 & 4170 &  0.03 & {\footnotesize 2}       & 3.03 & -7.13 & 0.4 & 0.2 \\ 
TW~Hya	  & 11:01:51.91	&-34:42:17.02  & K8 & 0.25 & 4000 & -1.92 & {\footnotesize 3}        & 1.0  & -9.38 & 0.7 & 7.0 \\ 
MWC~275   & 17:56:21:29	&-21:57:21.88  & A1 & 36.0 & 9500 &  0.40 & {\footnotesize 4}        & 2.2  & -7.12 & 2.3 & 5.0 \\ 
\enddata
\tablecomments{In column (5) we report the stellar luminosity in solar luminosities, in column (6) the stellar temperature in K, 
	in column (7) the accretion luminosity in $L_\sun$/yr, in column (9) the stellar radius in solar radii, in 
	column (10) the mass accretion rate in $M_\sun$/yr, in column (11) the stellar mass in solar masses and in 
	column (12) the stellar age in Myr.}
\tablerefs{(1) \cite{har98},  (2) \cite{nat06}, (3) \cite{M00}, (4) \cite{gar06}, (5) Kenyon and Hartmann (1995), 
	(6) Calvet \& Gullbring (1998), (7) Muzerolle et al. (1998)
}
\end{deluxetable}

\clearpage

\begin{deluxetable}{lcccll}
\tablecaption{Summary of CARMA continuum observations at 230~GHz \label{tab:obs_1mm}}
\tablewidth{0pt}
\tablehead{
\colhead{Object} & \colhead{Config.} & \colhead{Beam FWHM (\arcsec)} & \colhead{Beam PA (\arcdeg)} & \colhead{Date} &
\colhead{Phase Calibrators}
}
\startdata
CY~Tau   & C   & 1.05$\times$0.72 &  77 & 2007 Nov 12 & 3C111, 3C123              \\
DG~Tau   & C   & 0.87$\times$0.78 & -50 & 2007 Sep 30 & 3C111, 0530+135, 3C123    \\
 ...     & C   & 0.83$\times$0.64 & -72 & 2007 Oct 08 & 3C111, 0530+135, 3C123     \\
 ...     & B   & 0.43$\times$0.27 & -74 & 2007 Dec 14 & 3C111, 0530+135, 0510+180   \\	
DM~Tau   & C   & 0.82$\times$0.60 & -78 & 2007 Nov 05 & 3C111, 0530+135, 3C120      \\
DN~Tau 	 & C   & 0.80$\times$0.58 & -76 & 2007 Nov 05 & 3C111, 0530+135, 3C120      \\
DR~Tau   & C   & 0.92$\times$0.76 & -83 & 2007 Oct 09 & 3C111, 0530+135, 0449+113   \\
...      & C   & 0.84$\times$0.70 &  82 & 2007 Oct 24 & 0530+135, 0449+113          \\
...      & B   & 0.46$\times$0.34 &  39 & 2007 Feb 06 & 0530+135, 0449+113          \\
GO~Tau   & C   & 0.87$\times$0.65 &  88 & 2007 Nov 07 & 0530+135, 3C123             \\
LkCa15   & C   & 0.83$\times$0.70 &  70 & 2007 Oct 27 & 3C111, 0530+135, 3C123      \\
RY~Tau   & C   & 0.89$\times$0.74 & -61 & 2007 Oct 01 & 3C111, 3C123                \\
...      & C   & 1.14$\times$0.60 & -73 & 2007 Oct 22 & 3C111, 0530+135, 3C123      \\
UZ~Tau~E\tablenotemark{a} & C   & 0.82$\times$0.69 &  79 & 2007 Oct 27 & 3C111, 0530+135, 3C123      \\
GSS~39   & C   & 1.42$\times$0.85 & -6  & 2008 Apr 12 & 1625-254, 1733-130          \\
SR~24~S\tablenotemark{b}  & C   & 1.45$\times$0.91 & -6  & 2008 Apr 13 & 1625-254, 1733-130          \\
\enddata
\tablenotetext{a}{UZTau~W, the other component of the UZTau system, was 
	detected at 4$\sigma$ level with an integrated 1.3~mm flux of about 30 mJy
	(see Fig.~\ref{fig:cont})}
\tablenotetext{b}{The other component of the SR~24 binary system, SR~24~N \citep{p08}, 
	was not detected. }
\end{deluxetable}

\begin{deluxetable}{llllll}
\tablecaption{Properties of the 1.3 mm dust emission \label{tab:gau_fit}}
\tablewidth{0pt}
\tablehead{
\colhead{Object} & \colhead{Flux~(mJy)} & \colhead{Source size} & \colhead{$i$~(\arcdeg)} & \colhead{PA~(\arcdeg)} & \colhead{$R_g$~(AU)} \\  
\colhead{} & \colhead{} & \colhead{FWHM~(\arcsec)} & \colhead{} & \colhead{} & \colhead{} 
}
\startdata
CY~Tau   & 117$\pm$20 & 1.25$\times$0.60 & 61 & 150  & 230 \\ 
DG~Tau   & 317$\pm$28 & 0.52$\times$0.46 & 28 & 11   & 95  \\ 
DM~Tau   & 90$\pm$8   & 0.89$\times$0.82 & 22 & 20   & 160 \\ 
DN~Tau   & 93$\pm$8   & 0.68$\times$0.53 & 39 & 86   & 125 \\ 
DR~Tau   & 109$\pm$11 & 0.48$\times$0.39 & 36 & 108  & 90  \\ 
GO~Tau   & 57$\pm$8   & 0.86$\times$0.69 & 37 & 107  & 160 \\ 
LkCa15   & 119$\pm$15 & 1.39$\times$0.71 & 59 & 55   & 250 \\ 
RY~Tau   & 227$\pm$20 & 0.63$\times$0.38 & 53 & 23   & 115 \\ 
UZ~Tau~E & 126$\pm$12 & 0.88$\times$0.64 & 43 & 66   & 160 \\ 
GM~Aur   & 189$\pm$15 & 1.38$\times$1.08 & 45 & 106  & 270 \\
\hline
GSS~39   & 282$\pm$20 & 1.33$\times$0.87 & 49 & 116  & 260 \\ 
SR~24~S    & 197$\pm$17 & 1.64$\times$0.96 & 54 & 60  & 320 \\ 
\hline
TW~Hya   & 543$\pm$45 & 1.06$\times$1.04 & 12 & 89   & 76  \\
MWC~275  & 705$\pm$12 & 1.49$\times$1.12 & 41 & 135  & 250 \\
\enddata
\end{deluxetable}

\clearpage

\begin{deluxetable}{lll|ll|ll}
\tablecaption{Dust properties \label{tab:dust}}
\tablewidth{0pt}
\tablehead{
                 &                    &                  & \multicolumn{2}{|c|}{$a_{max}$=100mm}            & \multicolumn{2}{|c}{$q$=3}\\
\colhead{Object} & $\alpha$ & $\beta$ & $q$ & $k_{1.3}$\tablenotemark{a} & $a_{max}$\tablenotemark{b} & $k_{1.3}$\tablenotemark{a} 
}
\startdata
CY~Tau   & 2.6 & 0.7 & 3.5  & 0.6 & 1     & 3.9 \\
DG~Tau   & 2.3 & 0.5 & 3.1  & 0.2 & 22    & 0.39\\
DM~Tau   & 2.9 & 1.1 & 4.0  & 1.8 & 0.16  & 6.6 \\
DN~Tau   & 2.7 & 0.8 & 3.7  & 1.2 & 0.55  & 5.6 \\
DR~Tau   & 2.4 & 0.6 & 3.3  & 0.3 & 1.9   & 2.6 \\
GO~Tau   & 3.4 & 1.5 & 4.5  & 0.4 & 0.13  & 6.8 \\
LkCa15   & 3.5 & 1.7 & 4.7  & 0.3 & 0.13  & 6.8 \\
RY~Tau   & 2.5 & 0.7 & 3.5  & 0.6 & 1.0   & 3.9 \\
UZ~Tau~E & 2.6 & 0.7 & 3.5  & 0.6 & 1.0   & 3.9 \\
GM~Aur   & 3.1 & 1.3 & 4.2  & 1.1 & 0.14  & 6.7 \\
\hline
GSS~39   & 2.8 & 0.9 & 3.8  & 1.6 & 0.3   & 7.4 \\
SR~24    & 2.6 & 1.1 & 4.0  & 1.8 & 0.16  & 6.6 \\
\hline 
TW~Hya	 & 2.5 & 0.8 & 3.6  & 0.9 & 0.55  & 5.6 \\
MWC~275  & 2.9 & 1.0 & 3.9  & 1.9 & 0.185 & 7.0 \\
\enddata
\tablenotetext{a}{1.3~mm dust opacity is in cm$^2$ per gram of dust.} 
\tablenotetext{b}{dust size is in mm.}
\end{deluxetable}

\begin{deluxetable}{l|lllll|ll} 
\tablecaption{Model fitting results \label{tab:res}}
\tablewidth{0pt}
\tablehead{
\colhead{Object} & \colhead{$i$} & \colhead{PA} & \colhead{$R_t$} & \colhead{$\Sigma_t$} & \colhead{$\gamma$}
	& \colhead{$Log(M_d)$} & \colhead{$R_d$} \\
\colhead{} & (deg) & (deg) & (AU) & (g/cm$^2$) & &($M_\sun$) & (AU)  
}

\startdata
CY Tau	& 51$\pm$7        & 148$\pm$8 & 55$\pm$5     &  10$\pm$2        & -0.3$\pm$0.3        & -1.16 & 197 \\ 
DG~Tau  & 18$\pm$10       & 15$\pm$27 & 21$\pm$1     & 608$\pm$24       & -0.5$\pm$0.2        & -0.38 & 89  \\ 
DM~Tau	& 25$\pm$10       & 3$\pm$70  & 86$\pm$32    & 1.5$\pm$0.8      &  0.8$\pm$0.1        & -1.63 & 481 \\ 
DN~Tau  & 30$\pm$10       & 61$\pm$18 & 28$\pm$3     & 13$\pm$3         &  0.0$\pm$0.5        & -1.73 & 125 \\ 
DR~Tau  & 37$\pm$3        & 98$\pm$5  & 21$\pm$1     & 80$\pm$4         & -0.3$\pm$0.5        & -1.20 & 86  \\ 
GO~Tau	& 25$\pm$25       & 90$\pm$90                & 110$\pm$80   & 4$\pm$2          &  0.7$\pm$0.4        & -1.15 & 670 \\ 
LkCa15	& 58$\pm$4        &48$\pm$4   & 60$\pm$4     & 31$\pm$7         & -0.8$\pm$0.4        & -0.72 & 241 \\ 
RYTau	& 60$\pm$3        &25$\pm$3   & 25$\pm$1     & 58$\pm$4         & -0.1$\pm$0.4        & -1.19 & 112 \\ 
UZTauE	& 43$^{+10}_{-20}$ &70$\pm$5   & 43$\pm$10    & 12$\pm$5         &  0.8$\pm$0.4        & -1.32 & 260   \\ 
GM~Aur  & 51$\pm$2	  & 55$\pm$2  & 56$\pm$1     & 12$\pm$1         &  0.4$\pm$0.1        & -1.14 & 350  \\ 
\hline
GSS39	& 46$\pm$7        &111$\pm$7  & 66$\pm$10    & 4.7$\pm$1.6      &  0.5$\pm$0.2        & -1.36 & 390 \\ 
SR24	& 65$\pm$7        &48$\pm$4   & 20$\pm$4     & 50$\pm$10        &  0.1$\pm$0.3        & -1.43 & 120 \\ 
\hline
TW~Hya	& 11$\pm$2	  & 65$\pm$3  & 17.5$\pm$0.5 & 60$\pm$2         & -0.3$^{+0.1}_{-0.4}$ & -1.49 & 73  \\ 
MWC~275 & 51$\pm$2        & 21$\pm$4 & 85$\pm$3     & 2.7$\pm$1.2      &  0.3$\pm$0.1        & -1.41 & 520 \\ 
\enddata
\end{deluxetable}

\clearpage

\clearpage

\clearpage

\clearpage

\clearpage

\begin{figure*}
    \centering	
    \subfloat[]{%
	\label{fig:disk_a}
	\includegraphics[angle=0, width=\textwidth]{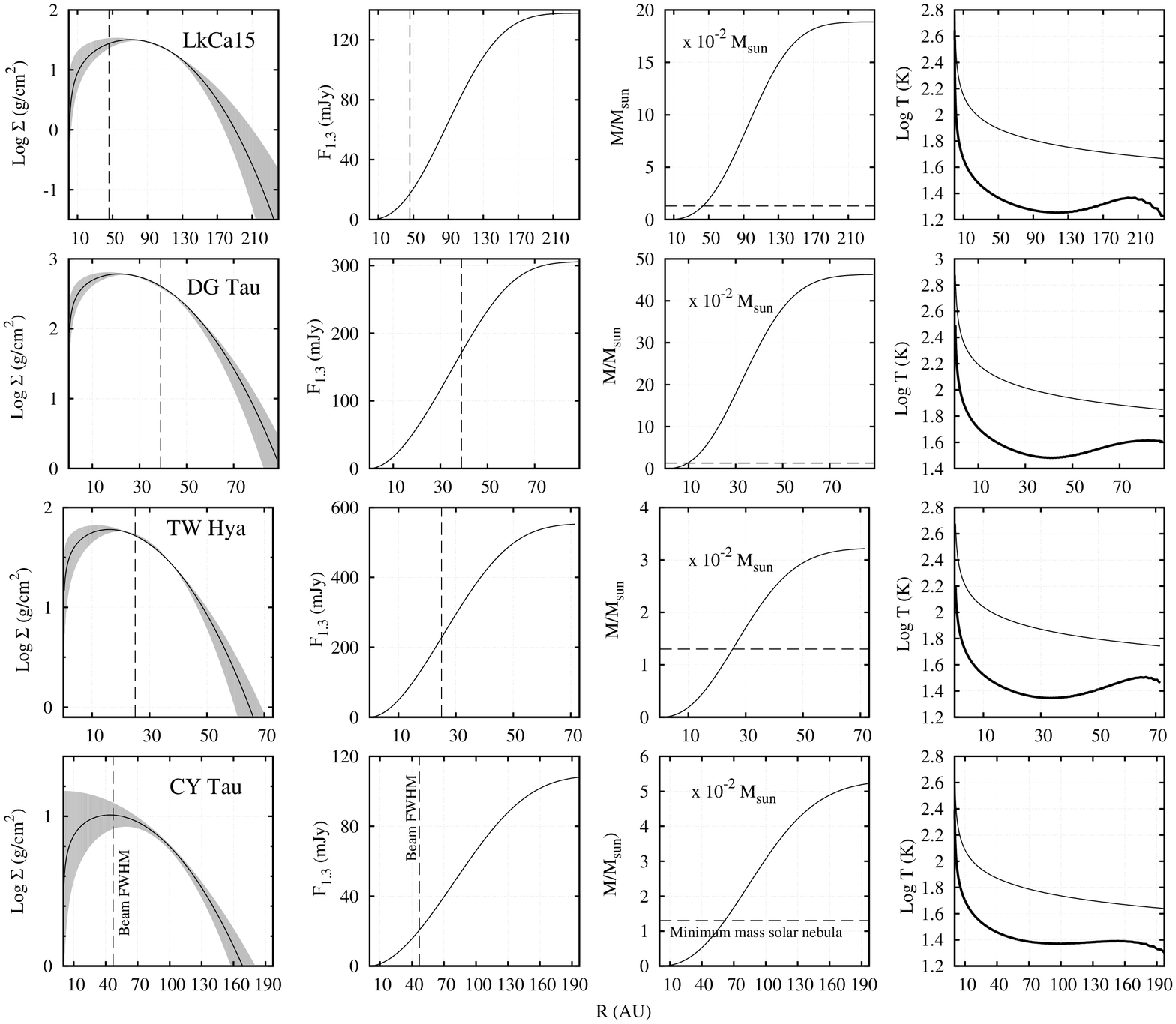}}
    \caption{ Radial profiles of the disk surface density, 
	cumulative flux at 1.3 mm, cumulative mass and disk temperature 
	for the best fit model from Table~\ref{tab:res}. The shaded 
	region in the surface density panels corresponds 
	to the 1$\sigma$ uncertainties obtained from MCMC fitting. The vertical 
	dashed lines indicate the spatial resolution of the observations.
	In the panels of column 3, the horizontal dashed line 
	at 0.02 $M_{\sun}$ is the minimum mass solar nebula \citep{hay81}, and
	the disk cumulative mass is expressed in units of $10^{-2}$ $M_\sun$.
	The disk interior temperature $T_i$ is represented by a thick line
	in the panels of column 4 while the disk surface layer temperature
	is a thin line.}
    \label{fig:disk}
\end{figure*}

\clearpage

\begin{figure*}
    \ContinuedFloat	
     \centering	
    \subfloat[]{%
	\label{fig:disk_b}
	\includegraphics[angle=0, width=\textwidth]{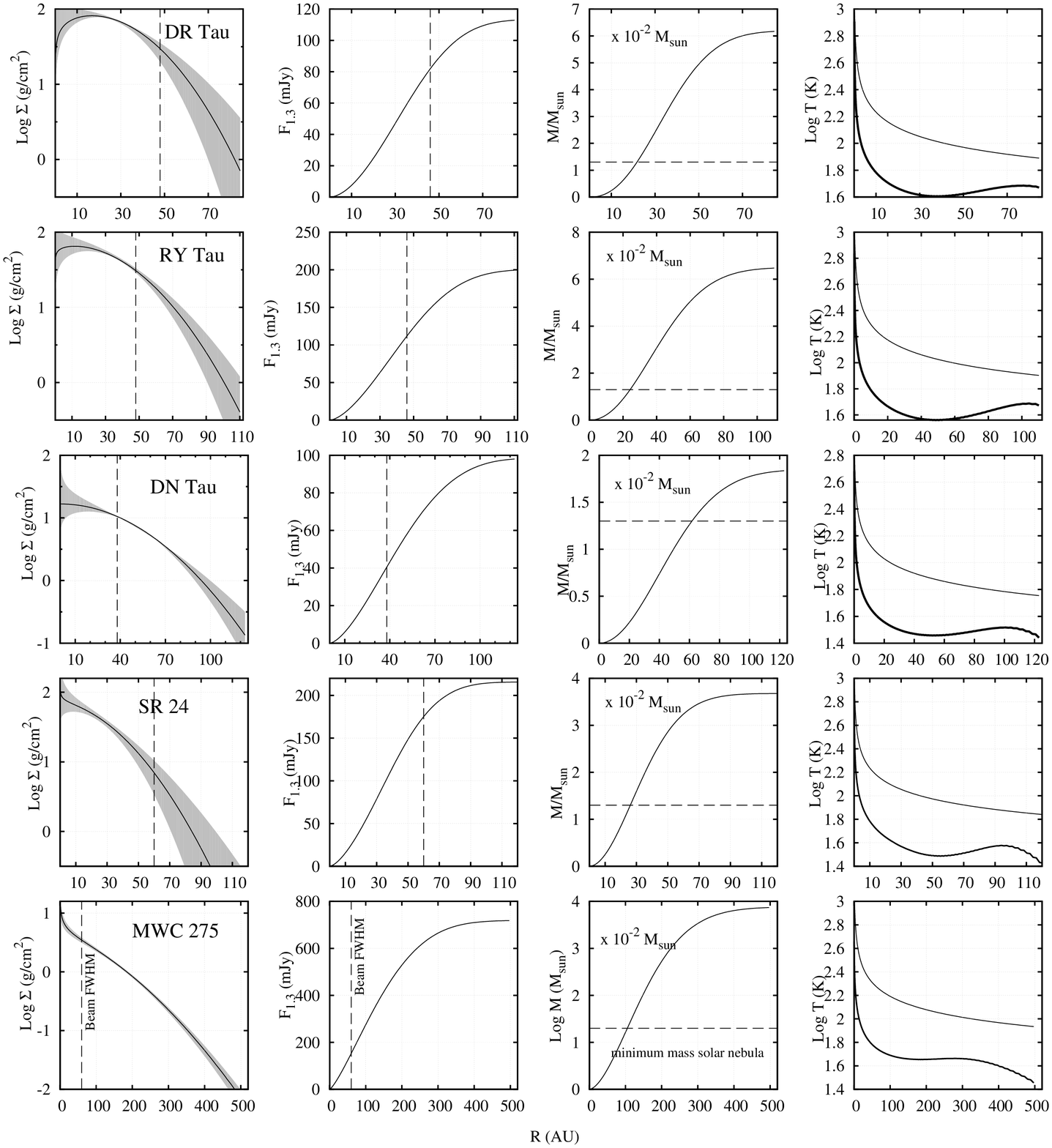}}
    \caption{continued}
\end{figure*}

\clearpage

\begin{figure*}
    \ContinuedFloat
    \centering	
    \subfloat[]{%
	\label{fig:disk_c}
	\includegraphics[angle=0, width=\textwidth]{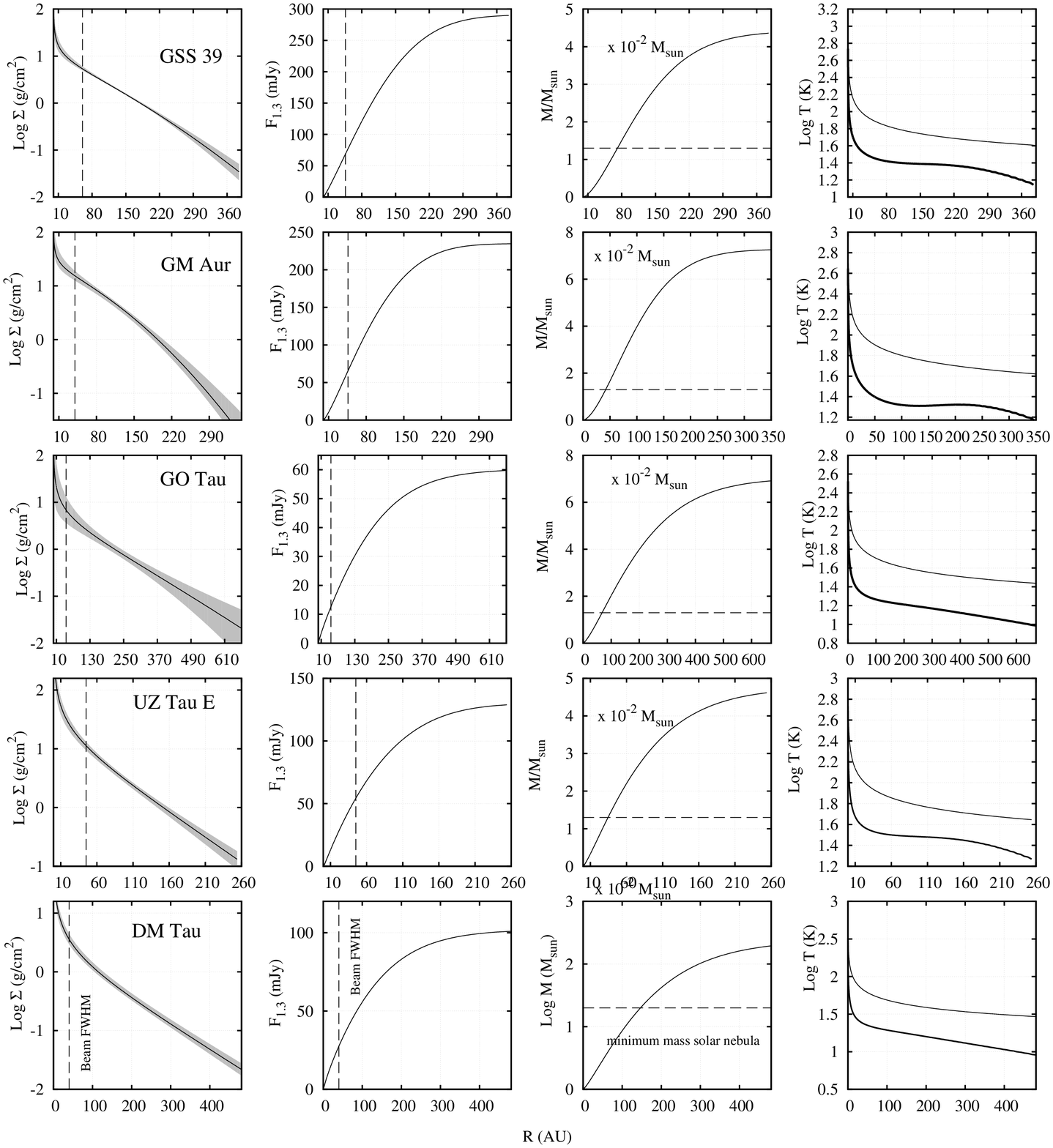}}
    \caption{continued}
\end{figure*}

\clearpage

\clearpage

\begin{figure*}
	\centering
	\subfloat[]{
		\label{fig:res_cont:a}
		\includegraphics[angle=0, width=\textwidth]{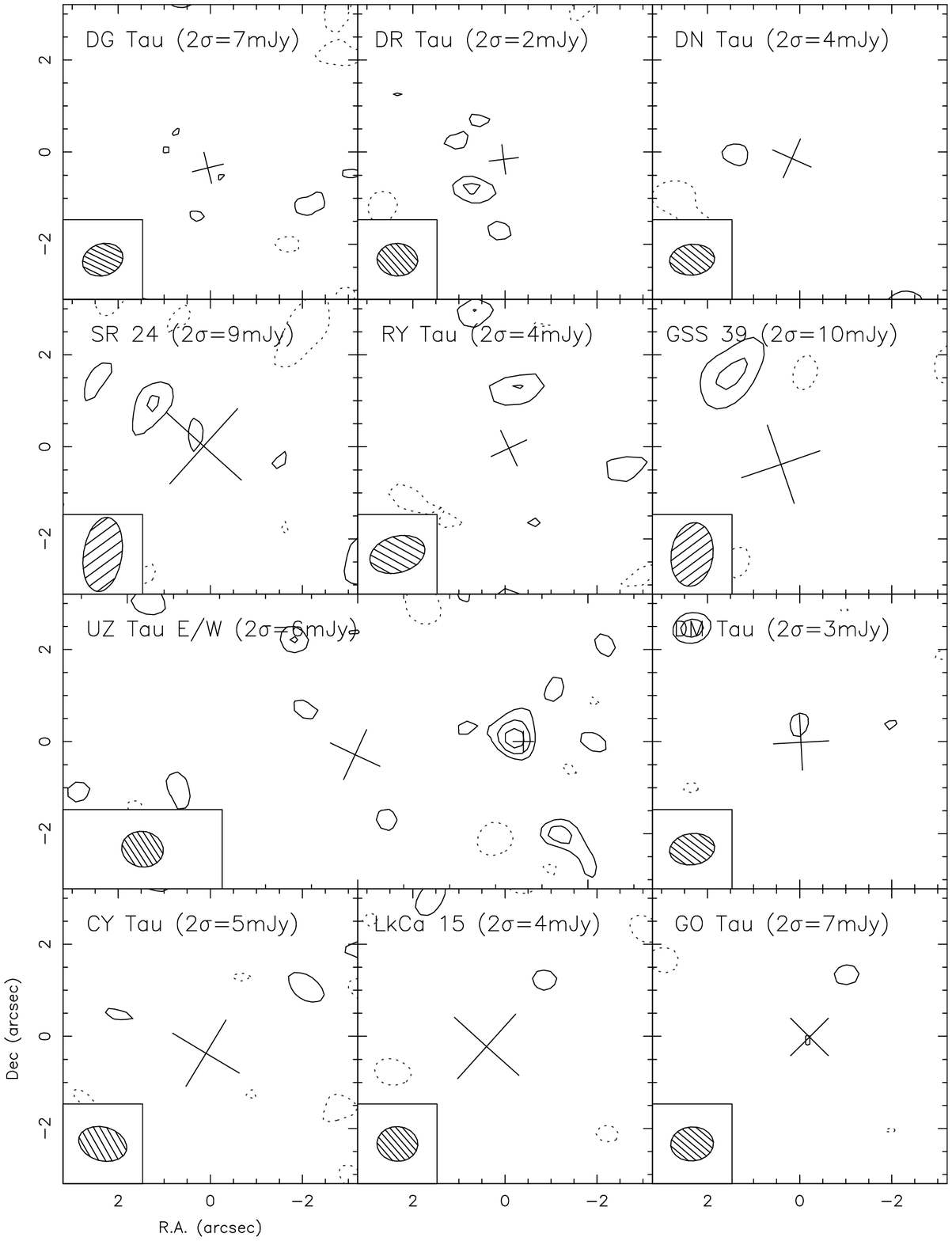}}
	\caption{Maps of the residuals calculated by subtracting the best fit 
	model from the observations. The contours
	start at the 2$\sigma$ level and are separated by 1$\sigma$. Cross 
	indicates the position of the source and the orientation of the disk.
	The smaller cross in the UZ~Tau~E panel indicates the position of UZ~Tau~W.}
	\label{fig:res_cont}
\end{figure*}

\begin{figure*}
	\ContinuedFloat
	\centering
	\subfloat[]{
		\label{fig:res_cont_b}
		\includegraphics[angle=270, width=\textwidth]{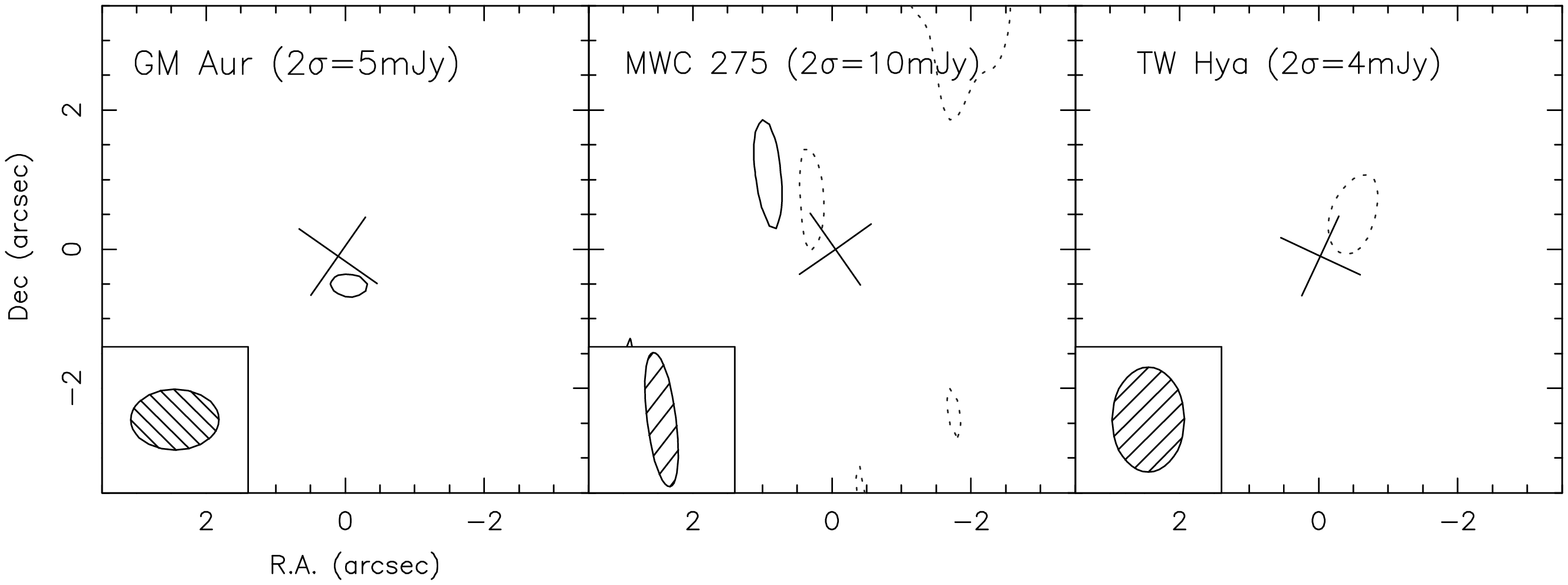}}
	\caption{continued}	
\end{figure*}

\clearpage

\clearpage

\clearpage

\clearpage

\clearpage


\clearpage

\clearpage

\clearpage

\begin{figure*}
	\centering
	\subfloat[]{
		\label{fig:prob_a}
		\includegraphics[width=\textwidth]{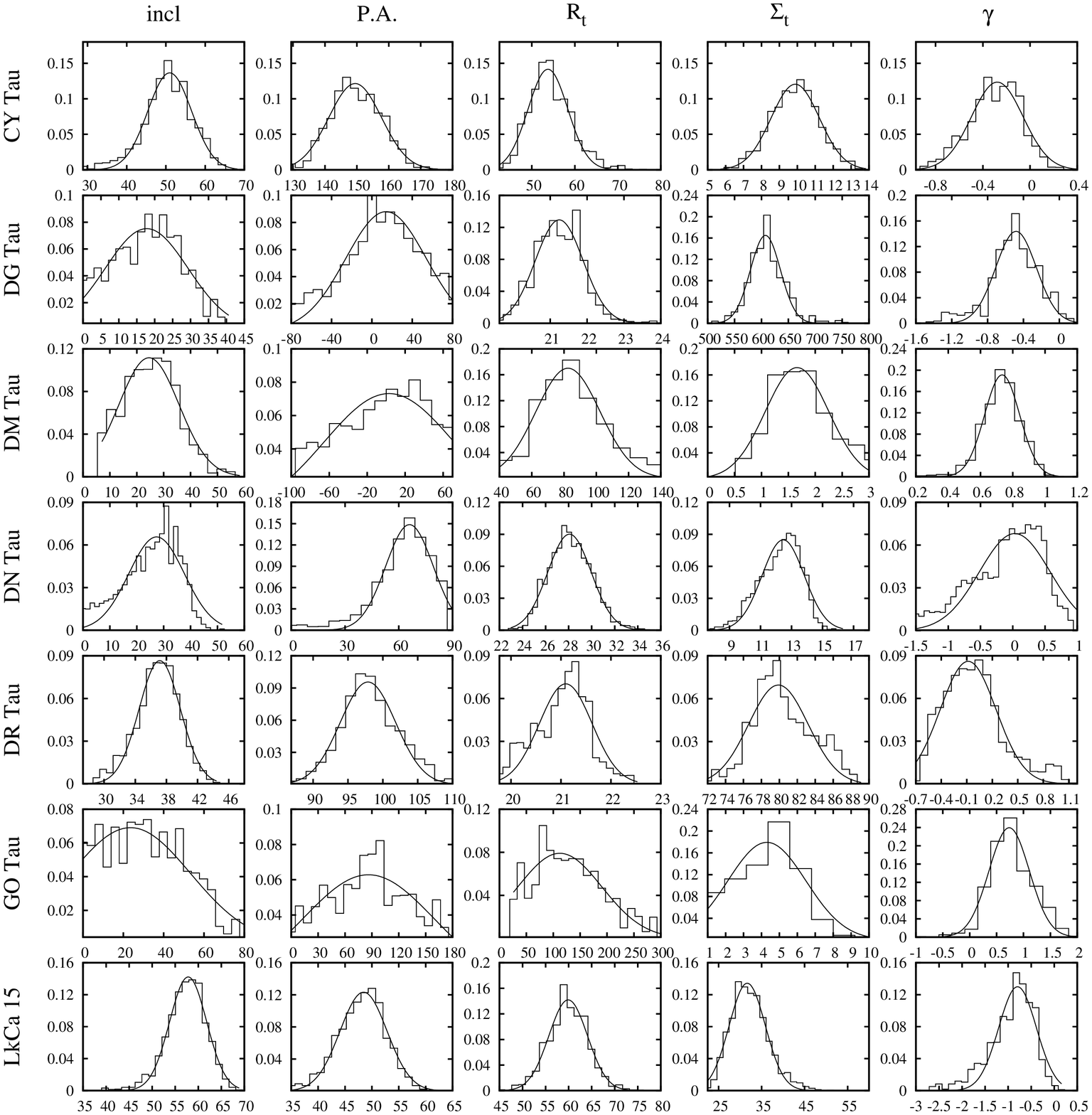}}
	\caption{Normalized probability distributions for the disk inclination 
	(\arcdeg), PA (\arcdeg), transition radius $R_t$ (AU), surface density 
	$\Sigma_t$ (g/cm$^2$) and  $\gamma$ obtained from
	the MCMC fitting process as discussed in Appendix~\ref{sec:MCMC}. 
	The solid line corresponds to the best fit Gaussian distribution 
	used to derive the parameter uncertainties reported in Tab.~\ref{tab:res}}
	\label{fig:prob}
\end{figure*}

\begin{figure*}
	\ContinuedFloat
	\centering	
	\subfloat[]{
		\label{fig:prob_b}
		\includegraphics[width=\textwidth]{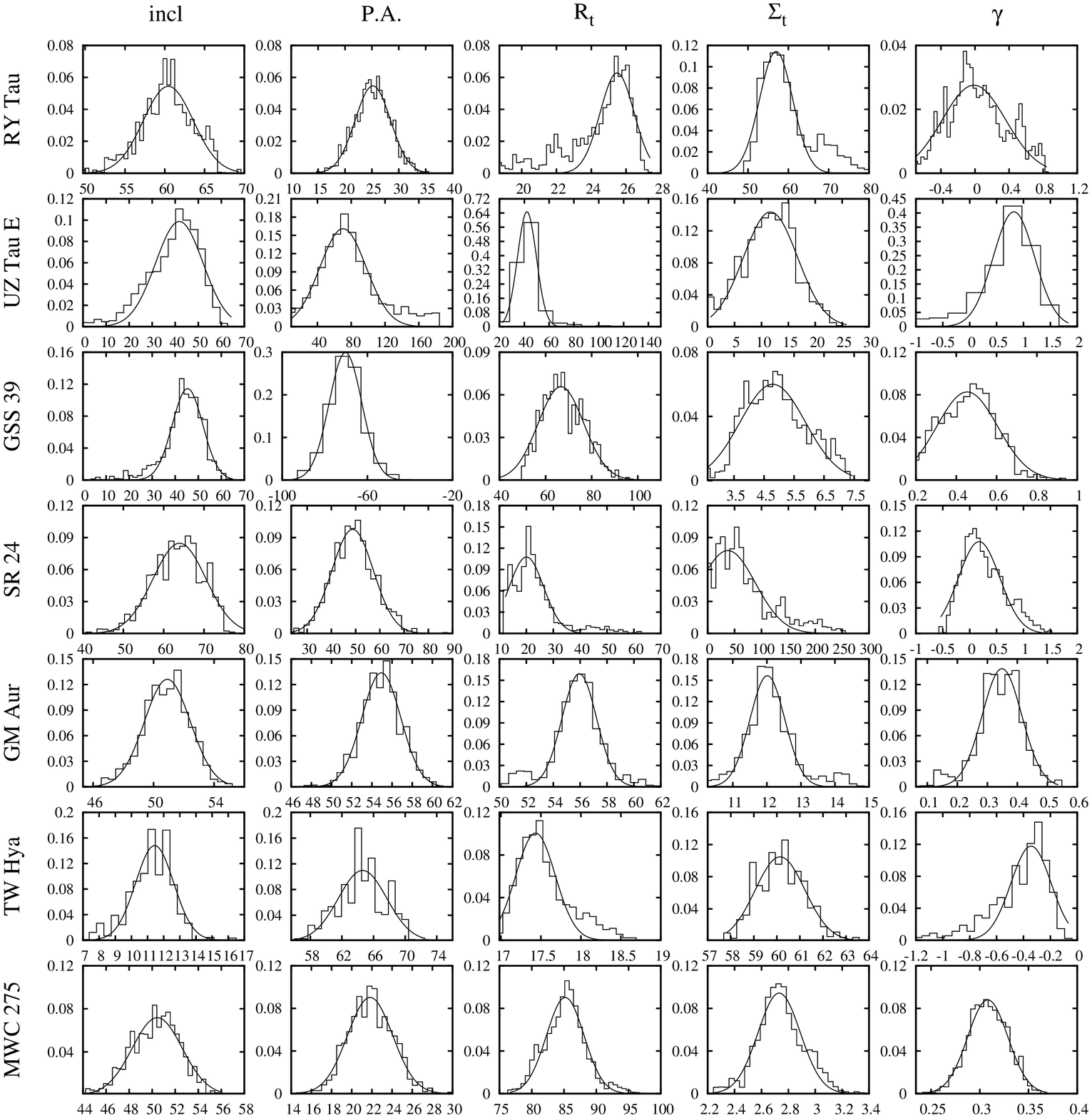}}
	\caption{continued}
\end{figure*}
	
\end{document}